\documentclass[structabstract]{aa}
\usepackage{graphicx}
\usepackage{txfonts}
\usepackage{amssymb}
\usepackage[below]{placeins}
\usepackage{natbib}
\usepackage{lscape}
\bibpunct{(}{)}{;}{a}{}{,}

\def \xa{XMMU\,J0302.2-0001}		% 066
\def \xb{XMMU\,J1532.2-0836} 		% 198
\def \xc{XMMU\,J0302.1-0000} 		% 064

\begin{document}

%________________________________________________________________

\title{Exploring the galaxy cluster-group transition regime at
high redshifts\thanks{Based on observations obtained with ESO Telescopes at the
Paranal Observatory under program ID 080.A-0659 and 081.A-0312, observations
collected at the Centro Astrn\'omico Hispano Alem\'an (CAHA) at Calar Alto,
Spain operated jointly by the Max-Planck Institut f\"ur Astronomie and the
Instituto de Astrof\'isica de Andaluc\'ia (CSIC). X-ray observations were
obtained by XMM-\emph{Newton}.}}

\subtitle{Physical properties of two newly detected $z>1$ systems}

\authorrunning{R. \v{S}uhada et al.}

\author{
R. \v{S}uhada \inst{1}\thanks{email: rsuhada@mpe.mpg.de}
\and
R. Fassbender \inst{1}
\and
A. Nastasi \inst{1}
\and
H. B\"ohringer \inst{1}
\and
A. de Hoon \inst{2}
\and
D. Pierini\thanks{Visiting astronomer at MPE.}
\and
J. S. Santos \inst{3}
\and
P. Rosati \inst{4}
\and
M. M\"uhlegger \inst{1}
\and
H. Quintana \inst{5}
\and
A. D. Schwope \inst{2}
\and
G. Lamer \inst{2}
\and
J. Kohnert \inst{2}
\and
G. W. Pratt \inst{6}
}

\institute{Max-Planck-Institut f\"ur extraterrestrische Physik (MPE),
Giessenbachstrasse~1, 85748 Garching, Germany
\and
Astrophysikalisches Institut Potsdam (AIP),
An der Sternwarte~16, 14482 Potsdam, Germany
\and
European Space Astronomy Centre (ESAC)/ESA, Madrid, Spain
\and
European Southern Observatory (ESO), Karl-Schwarzschild-Str.~2, 85748 Garching,
Germany
\and
Departamento de Astronom\'ia y Astrof\'isica, Pontificia Universidad
Cat\'olica de Chile, Casilla 306, Santiago 22, Chile
\and
CEA \/ Saclay, Service d'Astrophysique, L'Orme des Merisiers, B\^at. 709, 91191
Gif-sur-Yvette Cedex, France
}

\date{Received/accepted}
%________________________________________________________________

% \abstract{}{}{}{}{}
% 5 {} token are mandatory

  \abstract
   {Multi-wavelength surveys for clusters of galaxies are opening a window on the
elusive high-redshift ($z>1$) cluster population. Well controlled statistical
samples of distant clusters  will enable us to answer questions about
their cosmological context, early assembly phases and the thermodynamical evolution
of the intracluster medium.}
   {We report on the detection of two $z>1$
systems, \xa\ and \xb, as part of the XMM-\emph{Newton} Distant Cluster
 Project (XDCP) sample. We investigate the nature of the sources, measure their spectroscopic
 redshift and determine their basic physical parameters.}
   {The results of the present paper are based on the analysis of XMM-\emph{Newton}
   archival data, optical/near-infrared imaging and deep optical follow-up
   spectroscopy of the clusters.}
   {We confirm the X-ray source \xa\ as a gravitationally bound,
bona fide cluster of galaxies at spectroscopic
redshift $z=1.185$. We estimate its $M_{500}$ mass to $(1.6\pm0.3) \times 10^{14}$~M$_{\odot}$
from its measured X-ray luminosity. This ranks the cluster among intermediate mass system.
In the case of \xb\ we find the X-ray detection to be coincident with a dynamically bound
system of galaxies at $z=1.358$. Optical spectroscopy reveals the presence of a central
active galactic nucleus, which can be a dominant source of the detected X-ray emission
from this system. We provide upper limits of X-ray parameters for the system and discuss
cluster identification challenges in the high-redshift low-mass cluster regime.
A third, intermediate redshift ($z=0.647$) cluster, \xc, is serendipitously detected
in the same field as \xa. We provide its analysis as well.
}
 {}


   \maketitle
%
%________________________________________________________________

\section{Introduction}
\label{sec:intro}

%________________________________________________________________
\begin{table*}[th!]    % h = here ; positioning
\begin{center}
\caption{Observation log of the X-ray coverage of \xa\ and \xb. The quoted
exposure times are net clean times. MOS1 and MOS2 exposure times are averaged.
The off-axis angle for the XMM-\emph{Newton}
observations is an average angle from all three detectors.}
\label{tab:xobs}
\begin{tabular}{ c c c c c c }
\hline \hline
Cluster & Instrument & OBSID & Exp. time (PN/MOS) [ks] & Off-axis angle  \\
\hline
\xa\ & XMM & 0041170101 & 36.4/46.2 & $10.9\arcmin$ \\
\xb\ & XMM & 0100240701 & 13.2/17.4 & $5.0\arcmin$ \\
\xb\ & XMM & 0100240801 & $19.4/25.8^{\dag}$ & $6.1\arcmin$ \\
\hline
\end{tabular}
\end{center}
\footnotesize{$^{\dag}$ This observation is heavily contaminated by
quiescent proton flaring and used only for systematics check
(Sect.~\ref{sec:gca} and Sect.~\ref{sec:198discuss}).}
\end{table*}

\begin{table*}[th!]    % h = here ; positioning
\begin{center}
\caption{Summary of the optical/near-infrared observations of
\xa\ and \xb\ analysed in this work. The grism column applies
for spectroscopic observations, the band for imaging.
Exposure times and seeing are reported for each band individually.}
\label{tab:optobs}
\begin{tabular}{ c c c c l c c }
\hline \hline
Cluster & Data & Exposure Time  &  Prog. ID  &   Date  &   Grism/Band  &     Seeing \\
\hline
\xa & VLT/FORS2/ MXU spec. & 3h ($8\times1308$ sec) & 080.A-0659 & 2008 Jun 6    & 300 I & $1.03\arcsec-1.26\arcsec$ \\
\xb & VLT/FORS2/ MXU spec. & 3h ($8\times1308$ sec) & 081.A-0312 & 2008 Apr 4, 7 & 300 I & $0.79\arcsec-1.36\arcsec$ \\
\hline
\xa & CAHA 3.6m$^{\dag}$/Omega2000 imag. & 50 min/23 min  &  & 2006 Jan 3, 4    & H/z & $1.34\arcsec/1.49\arcsec$ \\
\xa & VLT/FORS2 imag. & 20 min & 079.A-0119 & 2007 Feb 23 & R & $0.7\arcsec$ \\
\xb & VLT/FORS2 imag. & 16 min/8 min & 078.A-0265 & 2008 Apr 4, 7 & R/z & $0.6\arcsec/0.8\arcsec$ \\
\hline
\end{tabular}
\end{center}
\footnotesize{$^{\dag}$ Calar Alto Observatory.}
\end{table*}
%________________________________________________________________

The number of known galaxy cluster detections at high redshift ($z\gtrsim1$)
is constantly growing (see Appendix~\ref{sec:hzhist}). Recently we have witnessed
the detection of the first
spectroscopically confirmed clusters at redshift $>1.6$ by \citet{papovich10}
and \citet{tanaka10}. This distance record has been however
soon overtaken by the cluster XMMU\,J105324.7+572348 with $z=1.753$
\citep{henry10}. Finally, \citet{gobat11} reported on the detection of
a remarkable structure which is consistent with a low mass cluster
at redshift 2.07. The nature of another potentially very distant system, JKCS041
detected by \citet{andreon09}
at a photometric redshift of 1.9, was contested by \citet{bielby10}. Recently obtained
deep z' and J imaging, however seems to confirm the presence
of a cluster at $z=2.2$ (based on a red sequence redshift estimation).

For the first time we can start constructing sizable
cluster samples at $z>1$ as a consequence of several important factors. First, there is progress in
cluster search methods, both classical such as X-ray
\citep[][Nastasi et al., in prep.]{mullis05, stanford06, henry10, fassbender11a} and
optical/mid-infrared surveys \citep{gladders05, stanford05, gobat11,
papovich10} as well as new selection methods like surveys utilising the
Sunyaev-Zel'dovich effect \citep[SZE,][]{planck11, williamson11, vanderlinde10, marriage10}.

The second essential prerequisite is the availability of deep
spectroscopic data, required to the confirm the cluster candidates as
genuine gravitationally bound systems and to estimate their redshifts. This typically
requires considerable effort and exposure times, and therefore many of
the current crop of distant clusters are the results of observational campaigns
spanning several years.

In addition, at $z>1.5$ we are also
nearing to the edge of capabilities of even the largest optical spectroscopic
instruments, since at these redshifts the 4~000~\AA-break (an important
feature to anchor the redshift of passive galaxies) is redshifted beyond
10~000~\AA, towards the tails of the sensitivity curves of current
spectrographs. Fortunately, near-infrared spectroscopy is able
to overcome this problem \citep[e.g.][]{tanaka10}. Existing
(e.g. MOIRCS on the Subaru telescope and LUCIFER at
the Large Binocular Telescope) and upcoming (KMOS at the VLT) instruments
will soon be able to provide confirmation for new high redshift clusters
with a much higher efficiency than the optical spectrographs.

Even though studying high redshift systems is a
truly daunting task, the effort is
rewarded by gaining a direct view of the earliest assembly epochs of the most
massive Dark Matter (DM) halos today, their gas content - the intracluster medium
(ICM) - and their galaxy populations. Since the properties of the cluster
 population are intrinsically
 connected to the underlying cosmology, they provide a sensitive test of the
cosmological parameters.

The high-mass end of the cluster mass function at high-redshift provides the
best leverage when constraining cosmological parameters through their effect
on the distribution and growth of the large scale structure. Since massive
distant clusters are rare, this regime can be effectively probed only by
surveys which are able to cover large sky areas (and hence large survey
volumes) such as the SZE surveys. The selection function of these surveys
is (almost) independent of redshifts and their sensitivity is limited to
very massive clusters with minimal mass $3-5 \times 10^{14}$~M$_{\odot}$
\citep{vanderlinde10,williamson11,marriage10}.

However, the majority of the cluster population lies below this mass
threshold. Therefore, if we want to understand the thermodynamical
evolution of the ICM and the evolution of the galaxy population,
we have to look at lower mass systems down to the group regime.
For the purpose of this paper we will consider the cluster-group
transition regime to be around $\sim1 \times 10^{14}$~M$_{\odot}$.
This threshold region can be probed by X-ray surveys, but it is
at the very limit of feasibility of contemporary X-ray surveys
\citep[joint X-ray and near-infrared detections can reach slightly
lower limiting masses, e.g.][]{finoguenov10}. Accessing
this threshold population will, however, allow us to directly
calibrate the mass scaling relations for less massive systems
and study potential mass-dependent effects on the evolution
of the galaxy population of the clusters.

Already a simple
consideration from the virial theorem \citep{kaiser86} predicts a tight link
between the ICM's properties (luminosity, temperature, gas mass) and the
total mass (i.e. including DM). These quantities are thus not only of interest
from the point of view of characterising the physical conditions of a given cluster,
but also as an important observational input for cosmological
studies. While the scaling relations of nearby clusters are fairly well known
\citep[e.g.][]{pratt09,arnaud05}, the evolution of these relations
is only starting to be explored at redshifts $z\gtrsim0.5-0.7$
\citep{vikhlinin09cccp2,pacaud07}. The $z>1$ regime is still practically unexplored.

The redshift range $z=1-2$ is a transition period also for the galaxy
population of clusters.
Local clusters exhibit typically well-defined red-sequences populated
by passively evolving early-type galaxies.
While similar red sequences are found also in some of the high redshift
clusters, e.g. the very massive cluster XMMU J2235.3-2557
(Strazzullo et al. 2010; Rosati et al. 2009), we are finding more and more
cases, where star-formation is still ongoing \citep[]
[]{fassbender11b,hayashi10, hilton10}.

Sizable, well controlled cluster samples at high redshift are thus important
to address many questions about the cluster population as a whole, but also
about the underlying cosmology. In this paper we provide first details on two new,
X-ray selected clusters at $z>1$. In Sect.~\ref{sec:data} we describe their
detection and follow-up observations (imaging and spectroscopy). Optical
properties are summarized in Sect.~\ref{sec:opt} and the X-ray analysis in
Sect.~\ref{sec:gca}. We discuss the results and draw conclusions in
Sect.~\ref{sec:discuss} and \ref{sec:conclusions}, respectively. The analysis of a third,
intermediate redshift cluster ($z=0.647$), serendipitously detected together
with \xa, is provided in Appendix~\ref{sec:064}.

Throughout the article, we adopt a $\Lambda$CDM cosmology with
$(\Omega_{\Lambda}, \Omega_{M}, H_0, w) = (0.7, 0.3, 70$ km s$^{-1}$ Mpc$^{-1},
-1)$.
Physical parameters are estimated within an aperture corresponding to a factor 500
overdensity with respect to the \emph{critical} density of the Universe at the cluster redshift.
All quoted magnitudes are given in the AB system.

\section{Observations and data analysis}
\label{sec:data}
The analysis of the presented clusters is based on archival, medium-deep
X-ray observations and optical/near-infrared data (both imaging and spectroscopic)
collected in a follow-up campaign. All the observations are summarized in Tables
~\ref{tab:xobs} and ~\ref{tab:optobs}.

\subsection{Initial X-ray detection with XMM-\emph{Newton}}
\label{sec:detect}
Both \xa\ and \xb\ were detected as extended X-ray sources in the framework
of the XMM-\emph{Newton} Distant Cluster Project (XDCP).
The XDCP consists of 470 XMM-\emph{Newton} archival fields with a total
non-overlapping area close to 80 deg$^2$. The data was obtained from the
XMM data archive.\footnote{\texttt{xmm.esac.esa.int/xsa/}}
The initial cluster detection was performed with the XMM-\emph{Newton}
science analysis system SAS v6.5 utilising a sliding box
detection and a maximum likelihood
source fitting.\footnote{SAS tasks \texttt{eboxdetect} and \texttt{emldetect}.}
 Details of the source detection pipeline can
be found in \citet{fassbender-thesis}.

For the purposes of this paper we re-analyzed the observations containing both
sources with the current updated version
of SAS (v10.0). The details of the observations are summarized in Table
~\ref{tab:xobs}.

\subsubsection*{\xa}
\xa\ was detected in the XMM-\emph{Newton} observation OBSID: 0041170101 with
39.3 ks PN exposure time and 50.4 ks in
either MOS camera. We identified and excised a time period strongly affected by
soft proton flaring in a two-step
cleaning process yielding 36.4 ks PN and 46.2 ks MOS clean exposures.
We find no residual quiescent soft proton contamination in any of the detectors.

The source is detected at the coordinates
$(\alpha, \delta) =  (03^{\mathrm{h}}\,02^{\mathrm{m}}\,11.9^{\mathrm{s}}, -00\degr\,01\arcmin\,\,\,34.3\arcsec)$ (J2000) at a relatively high off-axis angle of $11\arcmin$  with
very high detection and extent significance (both $\gtrsim10\sigma$).
The beta model core radius is $r_C=14.4\arcsec$, based on a fit with
 a fixed $\beta=2/3$. In an aperture of $1\arcmin$ we detected 130 source
counts in PN and 80 in the combined MOS detectors.

\subsubsection*{\xb}
The second X-ray source, \xb, is found in two XMM-\emph{Newton} observations (OBSID:
0100240701 and 0100240801) at coordinates $(\alpha, \delta) =  (15^{\mathrm{h}}\,32^{\mathrm{m}}\,13.0^{\mathrm{s}}$
$-08\degr\,36\arcmin\,\,\,56.9\arcsec)$.
The off-axis angles are in both observations similar, $\sim5\arcmin-6\arcmin$. Pointing 0100240801 is slightly
deeper with 19.4/25.7 ks clean time
 in PN/MOS compared to 13.2/17.3 ks of 0100240701. Unfortunately, after
inspecting the light curve of observation 0100240801
  we find  a steady decline of the count rate along the whole duration of the
observation - a clear indication of a
residual quiescent (i.e. non-flaring) soft proton contamination. To confirm this
suspicion we use a diagnostic
 test suggested by \citet{deluca04}. By looking at the count rate ratio inside
and outside the field of view of each
detector in the $8 - 10$~keV band, we find a $\sim50\%$ soft background
enhancement compared to the normal
level in PN and more than $90\%$ enhancement in both MOS cameras.
Observation 0100240701 is found to be completely uncontaminated.
In both observations we detect below 50 source counts, which is reflected
in the uncertainty of derived parameters.

The system was detected at a $\sim5\sigma$ significance level, however it was
classified as a point source in observation 0100240701.
It is only in the slightly deeper (but contaminated) observation, where the
source is flagged as extended with a $2\sigma$ significance and beta model core
radius of $r_C\simeq8\arcsec$. Therefore, in the source detection step the extent
is  established only tentatively. We describe an in-depth investigation of the
extent significance in Sect.~\ref{sec:198discuss}, where we conclude that the
currently available data is not sufficient to confirm the extended nature of
the source with any statistical significance.
All quoted values for \xb\ will come from the
analysis of the uncontaminated field 0100240701, unless noted otherwise.

\subsubsection*{\xc}
We also identified an additional cluster candidate in
the first XMM-\emph{Newton} observation (OBSID: 0041170101) roughly $2\arcmin$ from \xa.
We obtained spectroscopy for member galaxies for both clusters simultaneously
in the same FORS2 pointing. This allowed
us to confirm also this second source, \xc, as a genuine cluster of galaxies at
intermediate redshift $z\approx0.65$. In the following we will focus on the
two $z>1$ clusters and we provide details for \xc\ in Appendix~\ref{sec:064}.

\subsection{Optical/Near-infrared observations}
In addition to the archival X-ray data, we have obtained optical/near-infrared
imaging and deep optical spectroscopy for the clusters. In this section
we provide the details of the available data and its analysis.

\subsubsection{Follow-up imaging and optical properties}
\label{sec:imaging}
The optical/near-infrared imaging data used in this work is summarized in
Table~\ref{tab:optobs}. This data was obtained prior to spectroscopy to allow
pre-selection of the cluster candidates. Here we use it to investigate the
basic optical properties of the clusters' galaxy populations.

\subsubsection*{\xa}
In order to identify the optical counterparts of \xa, we carried out medium
deep H and z-band imaging data with the prime-focus wide-field (field-of-view of $15.4\arcmin$ on the side)
near-infrared OMEGA2000 camera \citep{bailer-jones00} at the 3.5m Calar Alto telescope.
The observations were performed on 3rd and 4th January (H and z band respectively) 2006
under clear conditions (calibration with on-chip 2MASS stars was done in photometric conditions).
We reduced the data with the designated OMEGA2000 NIR pipeline \citep{fassbender-thesis}.
The individually reduced frames are visually checked and co-added. The total exposure time
of the final stacked images is 50 min in H band (75 co-added frames) and 23 min in z band (23 frames).
We reach a $50\%$ completeness limit (AB) of H$_{lim} = 22.4$ mag and z$_{lim} = 23.5$ mag
with FWHM(H)$=1.34\arcsec$ and FWHM(z)$=1.49\arcsec$.

The photometry catalog was obtained by running \texttt{SExtractor} \citep{bertin96} in dual
image mode with the unsmoothed H-band image used as the detection image.
We then cross-checked the catalog with available SDSS photometry.

The VLT/FORS2 imaging (Prog. ID: 079.A-0119(A)) was carried out in
the R-band at a fairly good seeing of $\sim0.7\arcsec$ and photometric conditions. With
total clean exposure time of $20$~min, it is a valuable complement to the Calar
Alto imaging data. For the reduction of the pre-imaging data we followed
the same procedure as \citet{schwope10} and \citet{fassbender11a}.

\begin{figure*}[ht!]
\begin{center}
\includegraphics[width=0.48\textwidth]{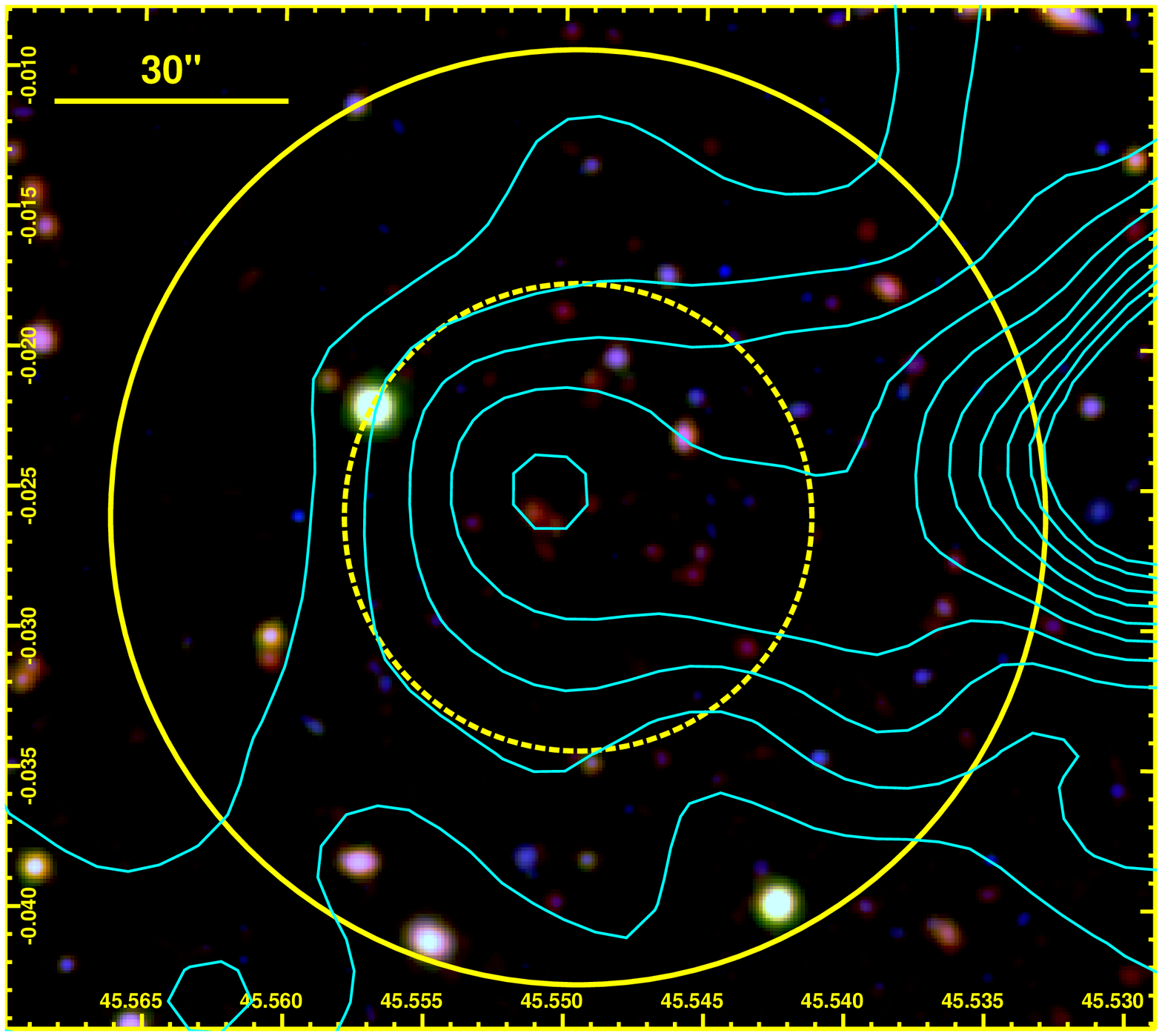}
\includegraphics[width=0.48\textwidth]{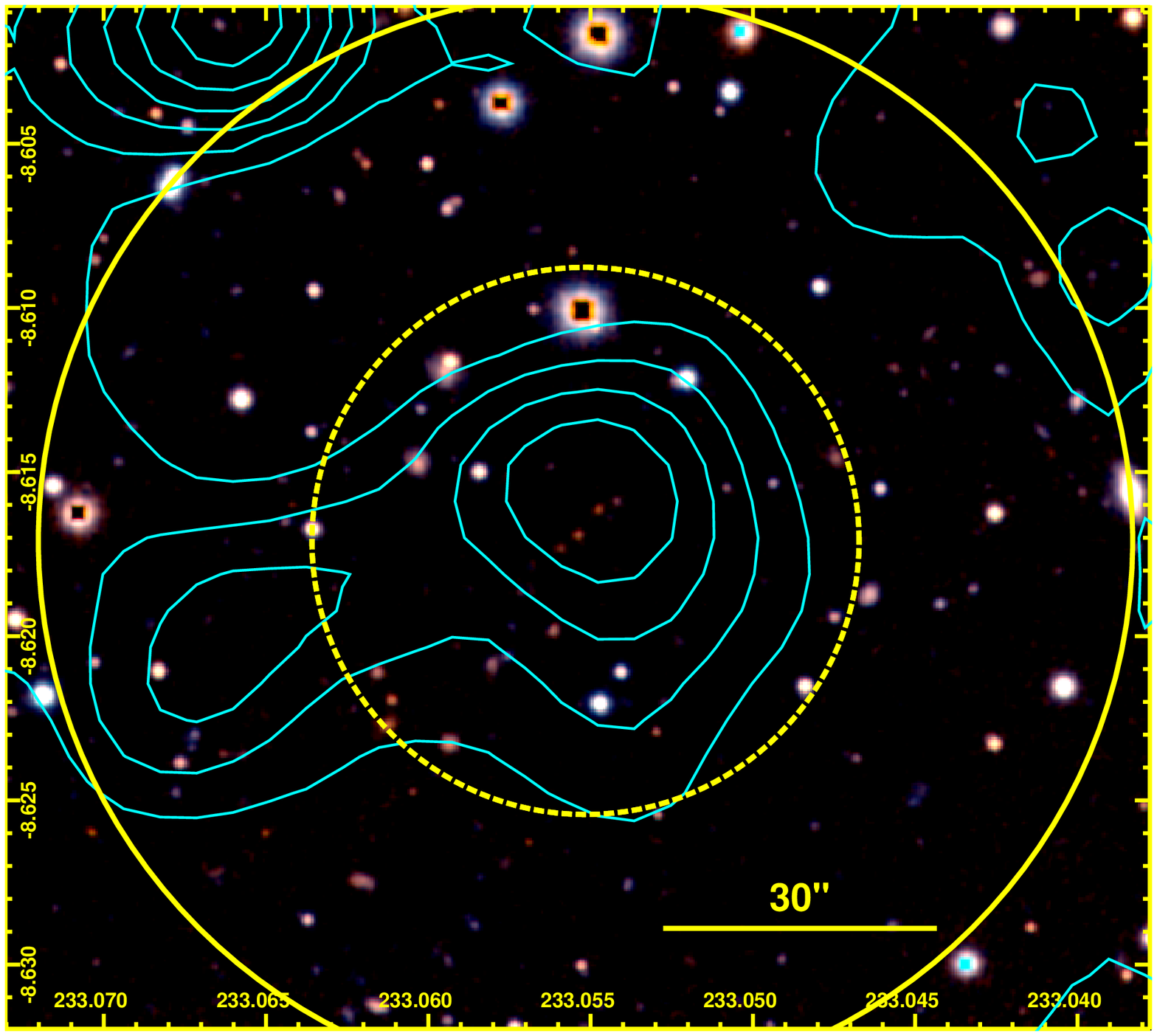}
\includegraphics[width=0.48\textwidth]{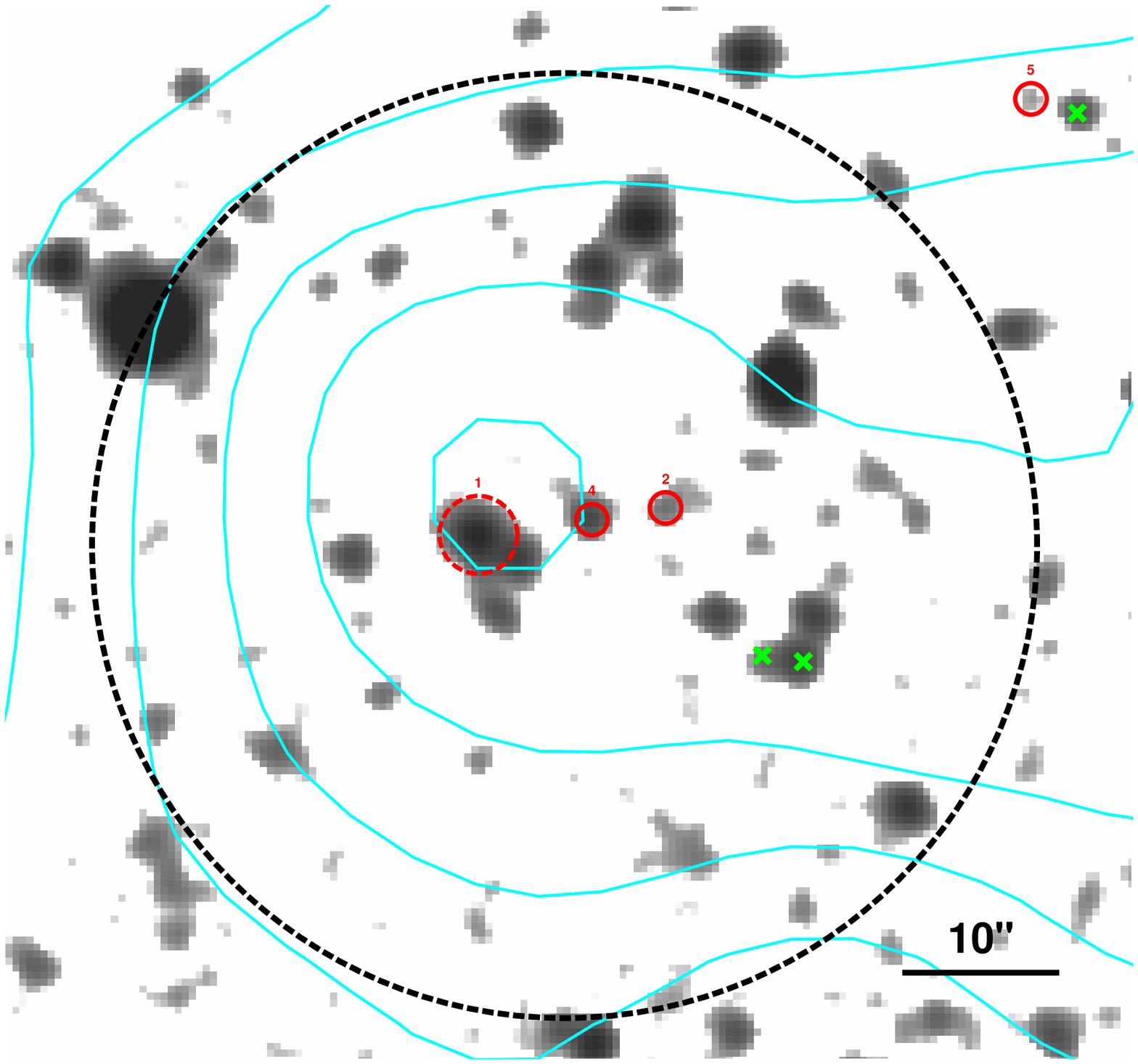}
\includegraphics[width=0.48\textwidth]{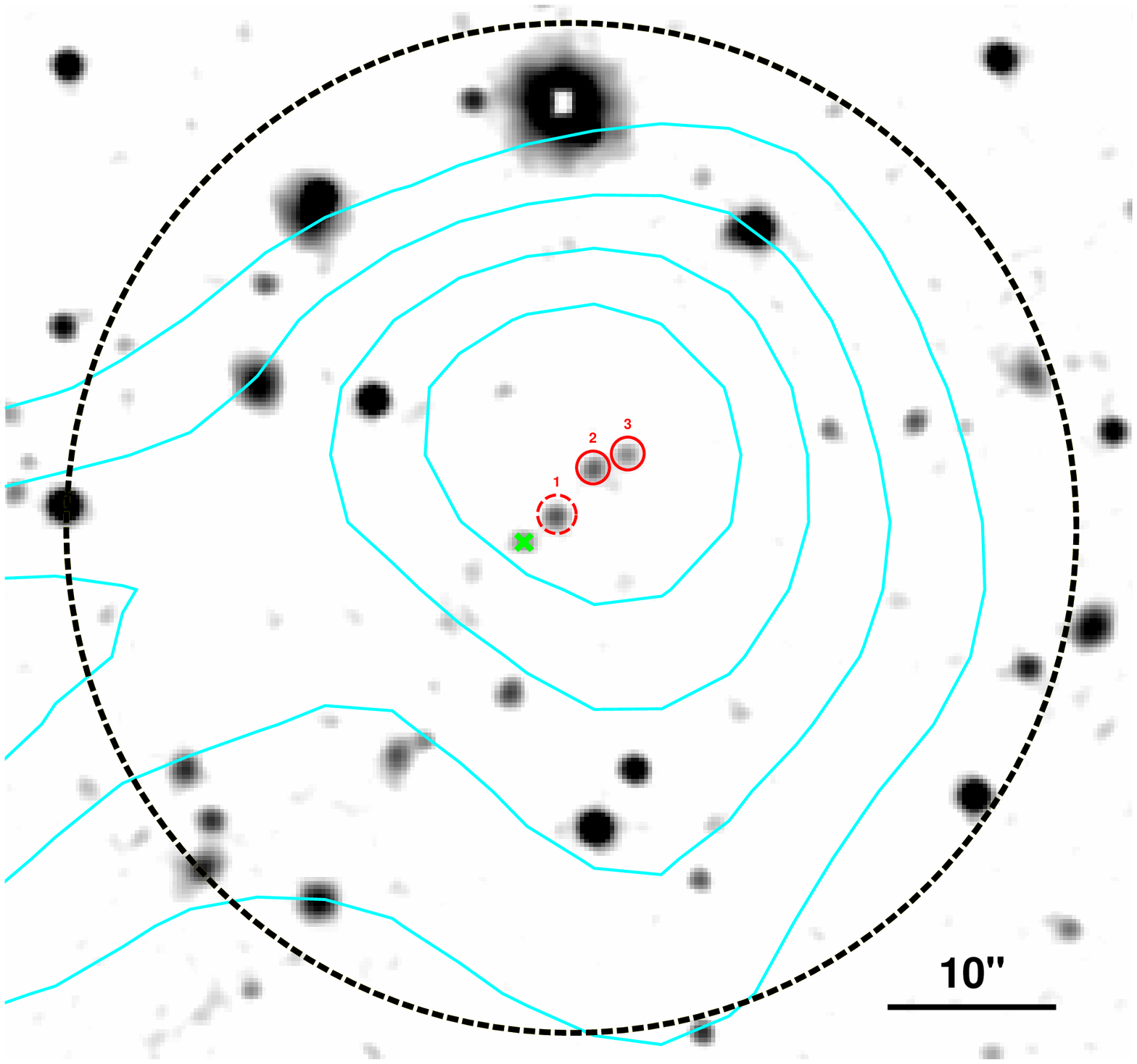}
\end{center}
\caption{Optical/Near-Infrared images of the clusters.
\emph{Top:} Pseudo-color image of the clusters \xa\ (left, red channel: H band,
green: z, blue: R) \xb\ (right, red channel: z,
green: z+R, blue: R). Adaptively smoothed X-ray contours are overlaid in cyan.
Solid/dashed circles mark a $60\arcsec$/$30\arcsec$ radius region centered on
the X-ray detection. An associated overdensity of red galaxies is apparent
in both cases. \emph{Bottom:} A high contrast zoom on the central regions of
the clusters (dashed circle has a $30\arcsec$ radius). \xa\ is displayed in the
H band whereas \xb\ in the z band. Red circles mark the confirmed spectroscopic
members with properties listed in Table~\ref{tab:gals} and spectra displayed in
Fig.~\ref{fig:spec}. The BCG is marked with a dashed red circle (ID 1 for both
clusters). Spectroscopically confirmed foreground galaxies are indicated by green
crosses.}
\label{fig:opt}
\end{figure*}

In Fig.~\ref{fig:opt} (top left) we display a pseudo-color
image of \xa\ in the H/z/R bands (red/green/blue). A population of red
galaxies ($1.45<\mathrm{z}-\mathrm{H}\leq2.15$) is found to be coincident with the X-ray source. We show the
z$-$H vs. H color-magnitude diagram (CMD) of \xa\ in
Fig.~\ref{fig:cmd} (top). We also overplot the synthetic z$-$H color of a
Simple Stellar Population (SSP)
model (formation redshift z$_f$=5, solar metallicity) for the
cluster's redshift (red dashed line).
We find around 10 red galaxies within $30\arcsec$ from the X-ray center with
colors well matching the model prediction. The overdensity of red
galaxies compared to the field is at a $\sim25\sigma$ significance level - one
of the largest known overdensities at $z>1$.

The brightest cluster galaxy (BCG) is coincident with the X-ray emission peak and
 seems to be
undergoing merging activity (see Fig.~\ref{fig:opt}, bottom left). The very
bright blue object ($\mathrm{H}\approx18.4$~mag) at the cluster redshift (but beyond $30\arcsec$ from
its center) is an AGN with redshift
from SDSS (Sect.~\ref{sec:ned}). Galaxy ID: 6 (in Table~\ref{tab:gals})
also has a bluer color, which is consistent with the presence of a
very strong [O\,\textsc{ii}] emission line (see Fig.~\ref{fig:spec}).

\subsubsection*{\xb}
\begin{figure}[t]
\begin{center}
\includegraphics[width=0.48\textwidth]{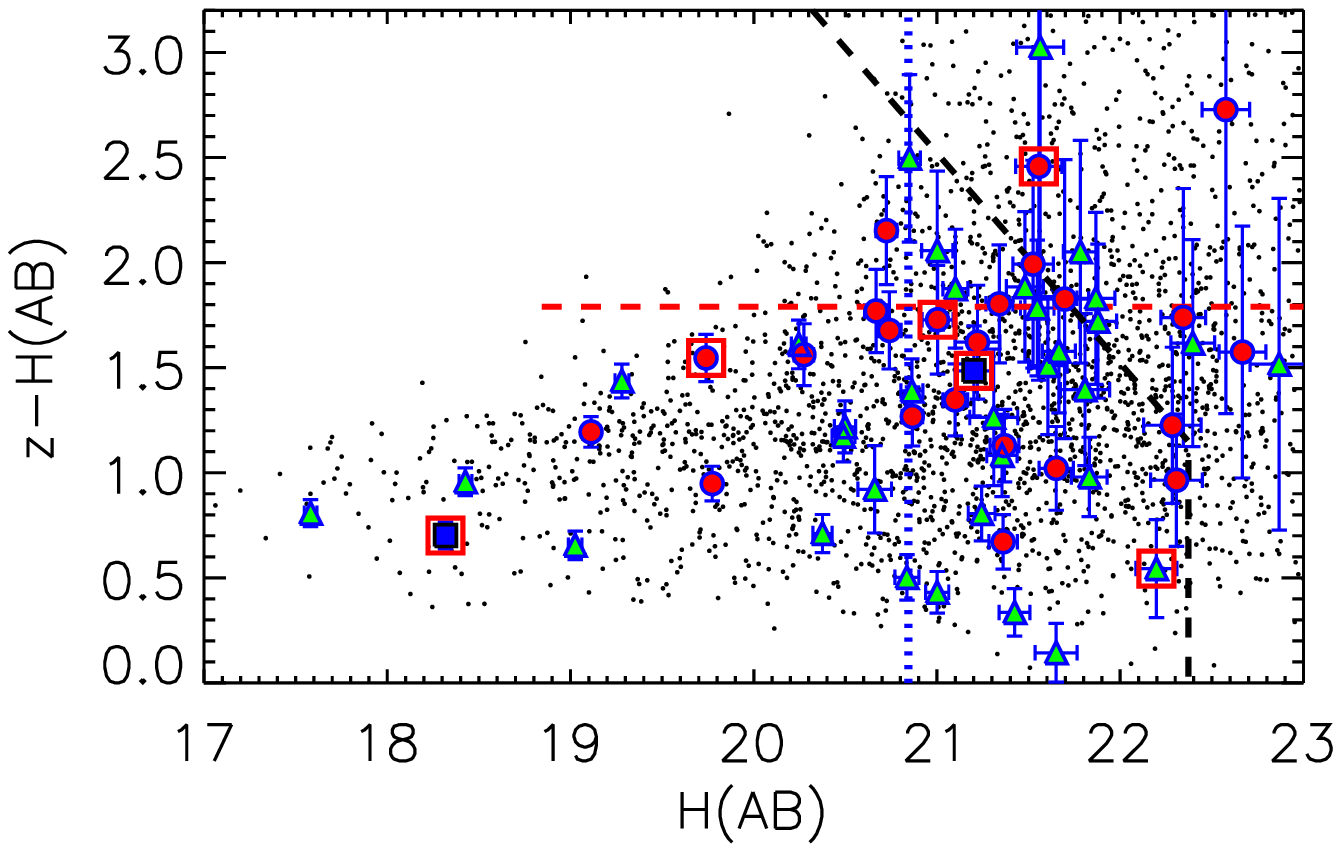}
\includegraphics[width=0.48\textwidth]{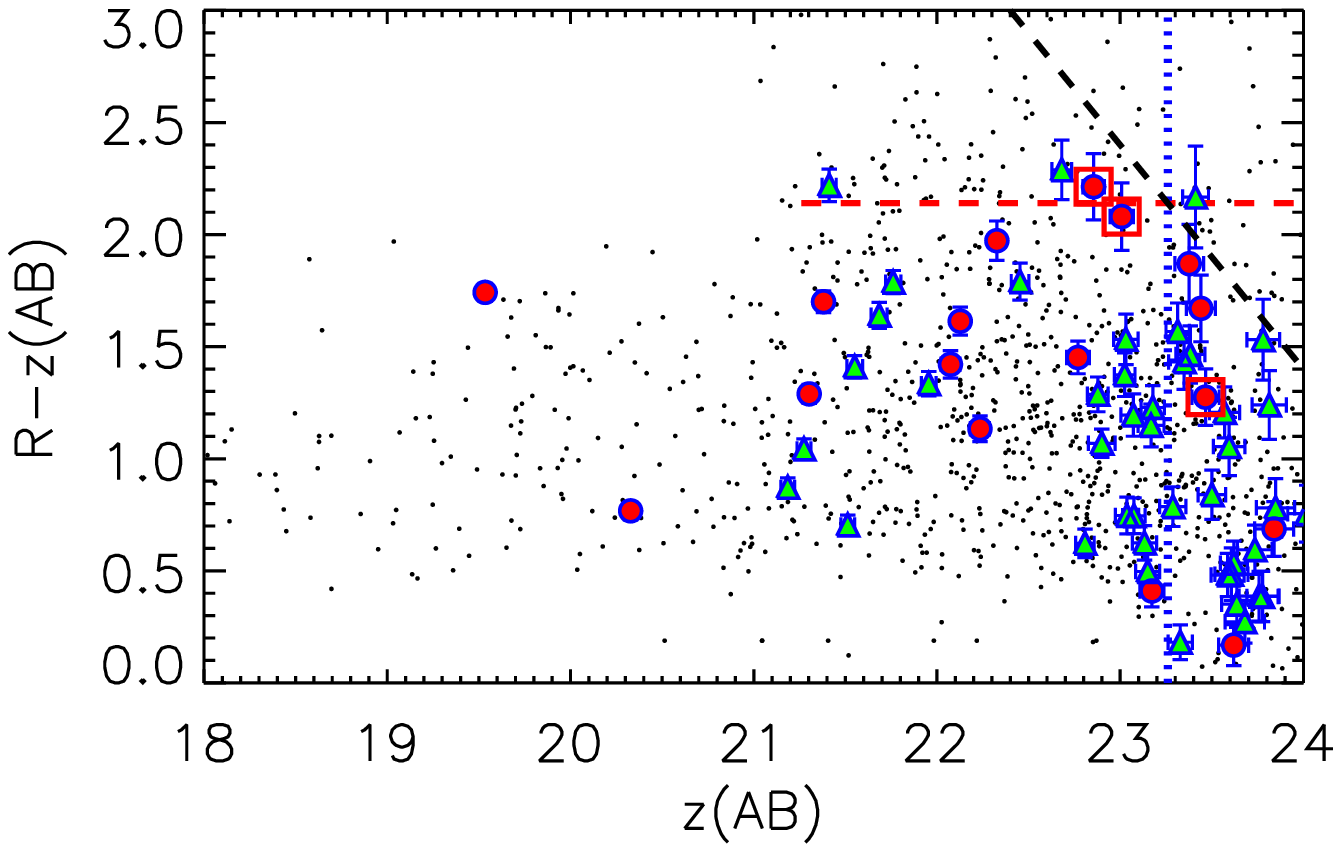}
\end{center}
\caption{\emph{Top:} The z-H vs. H color-magnitude diagram (CMD) for \xa.
Square boxes mark secure spectroscopic cluster members. Galaxies with
projected cluster-centric distances less than $30\arcsec$ are shown as
red circles, those with distances between $30\arcsec-60\arcsec$ as green triangles.
Galaxies with concordant redshift at $>60\arcsec$ distances have blue squares.
The
dashed black line marks the magnitude limits.
The apparent H band magnitude of an L$^{*}$ galaxy at the cluster redshift
is shown with a vertical blue dotted line. To help to guide the eye
we overplot the color of a solar
metallicity SSP model for the clusters' redshifts with formation
redshift z$_f=5$ (horizontal red dashed line).
\emph{Bottom:} The R-z vs. z CMD of \xb. The symbols and colors
have the same meaning as in the above plot.
Two of the spectroscopic members
lay on the model prediction for a red sequence at this redshift, the third
member has a significantly bluer color.}
\label{fig:cmd}
\end{figure}

The imaging in the case of \xb\ consists of R and z band imaging obtained
 with the VLT/FORS2 instrument (Prog. ID: 078.A-0265) and seeing of $0.6\arcsec$
 and $0.8\arcsec$, respectively. The total exposure
time is 16 min in R band and 8 min in z.
The final R-z vs. z color-magnitude diagram of \xb\ is displayed in
Fig.~\ref{fig:cmd} (bottom). The dashed red line shows the R$-$z
color of a spectro-photometric sequence
(SSP model, z$_f$=5, solar metallicity) at the redshift of the cluster.
Two spectroscopic members have colors consistent with this simple model.

As can be noted from the figure,
these two galaxies are very close to the completeness limit of our data and we
can thus see only the very brightest end of the galaxy population. The third
galaxy (ID: 3) has bluer colors and a strong [O\,\textsc{ii}] emission line.
We also detect [Ne\,\textsc{iii}]$\lambda3869~\AA$ and
[Ne\,\textsc{v}]$\lambda3426~\AA$ emission lines.
It is therefore likely
that this galaxy harbours an obscured AGN (see Sect.~\ref{sec:198discuss}).

We have designated the brightest spectroscopic member as the cluster
candidate's BCG (ID: 1), with $\sim5.5\arcsec$ distance from the cluster center.
This galaxy is relatively faint (z$^{*}-0.4$). The brightest
galaxy lying exactly at the predicted SSP color is a spectroscopically confirmed
foreground galaxy. However, there is one galaxy slightly brighter than
the marked BCG within $30\arcsec$ from the X-ray center - which could also
be a BCG candidate. Unfortunately, we do not have spectroscopy for this source.
Compared to galaxy ID: 1 it has a slightly bluer color than the SSP prediction and a
slightly larger cluster-centric distance, i.e. ID: 1 still remains
the better BCG candidate.

\label{sec:opt}

\begin{table*}[ht]    % h = here ; positioning
\begin{center}
\caption{Spectroscopic redshifts of cluster member galaxies. The last column
lists the main detected spectral features. Forbidden lines are detected in emission,
the rest in absorption. The distance from the X-ray center in arcseconds
is given as the d$_{X}$ parameter.}
\label{tab:gals}
\centering
\begin{tabular}{l c c c c l}
\hline
\hline
  \multicolumn{1}{c}{ID} &
  \multicolumn{1}{c}{$\alpha$ (J2000)} &
  \multicolumn{1}{c}{$\delta$ (J2000)} &
  \multicolumn{1}{c}{z$_{spec}$} &
  \multicolumn{1}{c}{d$_{X}$ [$\arcsec$]} &
  \multicolumn{1}{c}{Features} \\
  \hline
  \textbf{\xa} \\
  \hline
  1 (BCG) & 03:02:12.260 & -00:01:33.87 & $1.1848 \pm 0.0007$ & 5.4   & [O\,\textsc{ii}]$^{\dag}$, Ca-K, 4\,000 \AA~break  \\
  2       & 03:02:11.462 & -00:01:32.01 & $1.1735    \pm 0.0007$ & 7.0   & Mg\,\textsc{ii},  Ca-H/K, 4\,000 \AA~break        \\
  3       & 03:02:16.181 & -00:03:32.28 & $1.1806    \pm 0.0004$ & 134.3 & [O\,\textsc{ii}], Ca-H/K$^{\dag}$                  \\
  4       & 03:02:11.774 & -00:01:32.61 & $1.1968    \pm 0.0007$ & 2.5   & Mg\,\textsc{ii}, [O\,\textsc{ii}], Ca-H, G band                      \\
  5       & 03:02:09.930 & -00:01:05.91 & $1.2042    \pm 0.0004$ & 41.0  & [O\,\textsc{ii}]$^{\dag}$                         \\
  6       & 03:02:15.228 & -00:01:49.87 & $1.1596    \pm 0.0004$ & 52.3  & [O\,\textsc{ii}] \\
  \hline
  \textbf{\xb} \\
  \hline
  1 (BCG) & 15:32:13.294 & -08:37:00.75 & $1.3592 \pm 0.0016$ & 5.5  & Mg\,\textsc{ii}, Ca-H/K                            \\
  2       & 15:32:13.149 & -08:36:57.97 & $1.3580 \pm 0.0007$ & 2.1  & Fe\,\textsc{ii}, Mg\,\textsc{ii}, Ca-H/K                      \\
  3       & 15:32:13.010 & -08:36:57.14 & $1.3568 \pm 0.0005$ & 0.4  & Fe\,\textsc{ii}, [O\,\textsc{ii}], [Ne\,\textsc{iii}], [Ne\,\textsc{v}]                    \\
\hline
\end{tabular}
\end{center}
\footnotesize{$^{\dag}$ The feature is faint.}
\end{table*}

\subsubsection{Spectroscopic confirmation}
\label{sec:spec}

\begin{figure*}[t!]
\begin{center}
\includegraphics[width=0.48\textwidth]{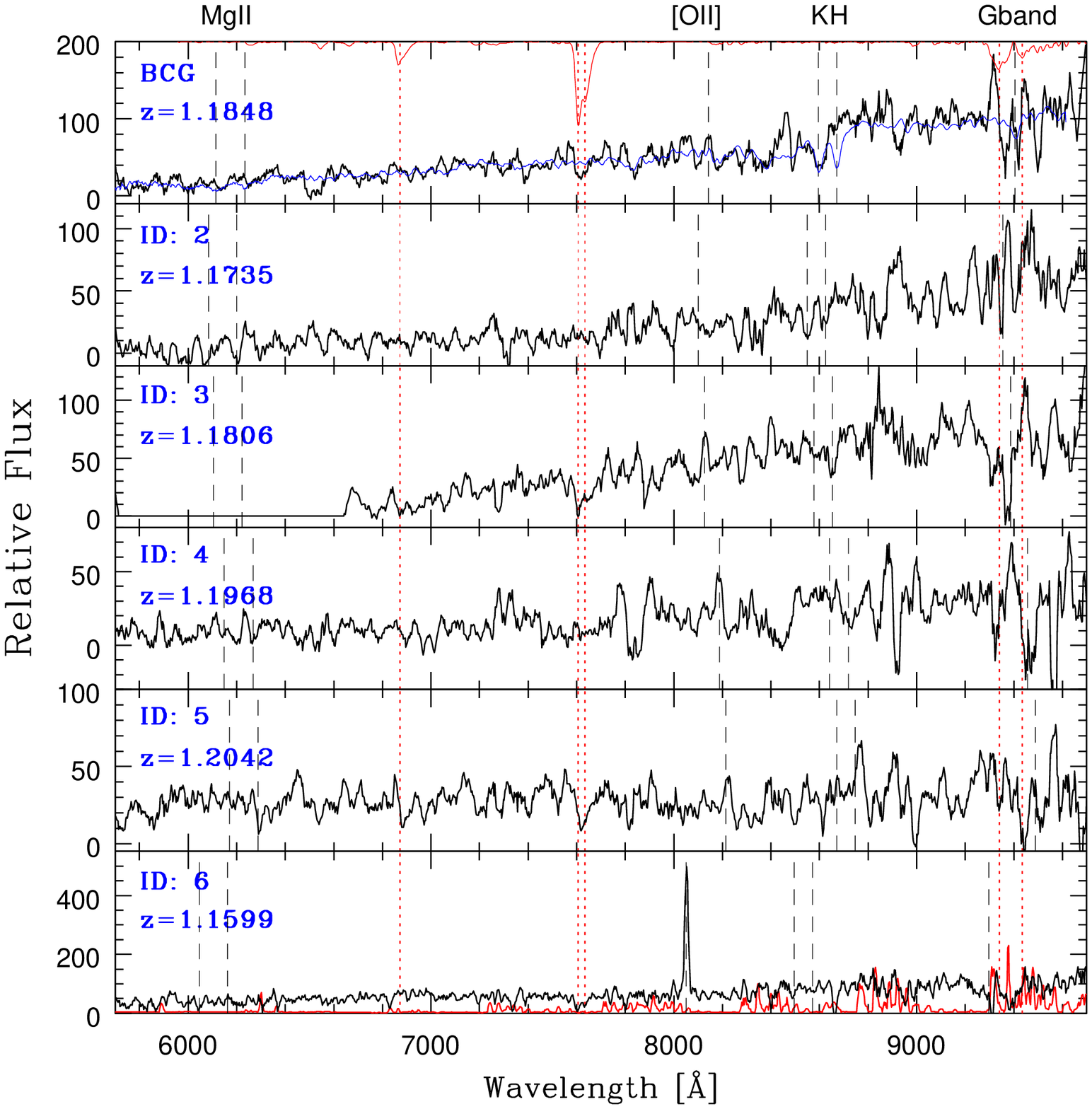}
\includegraphics[width=0.48\textwidth]{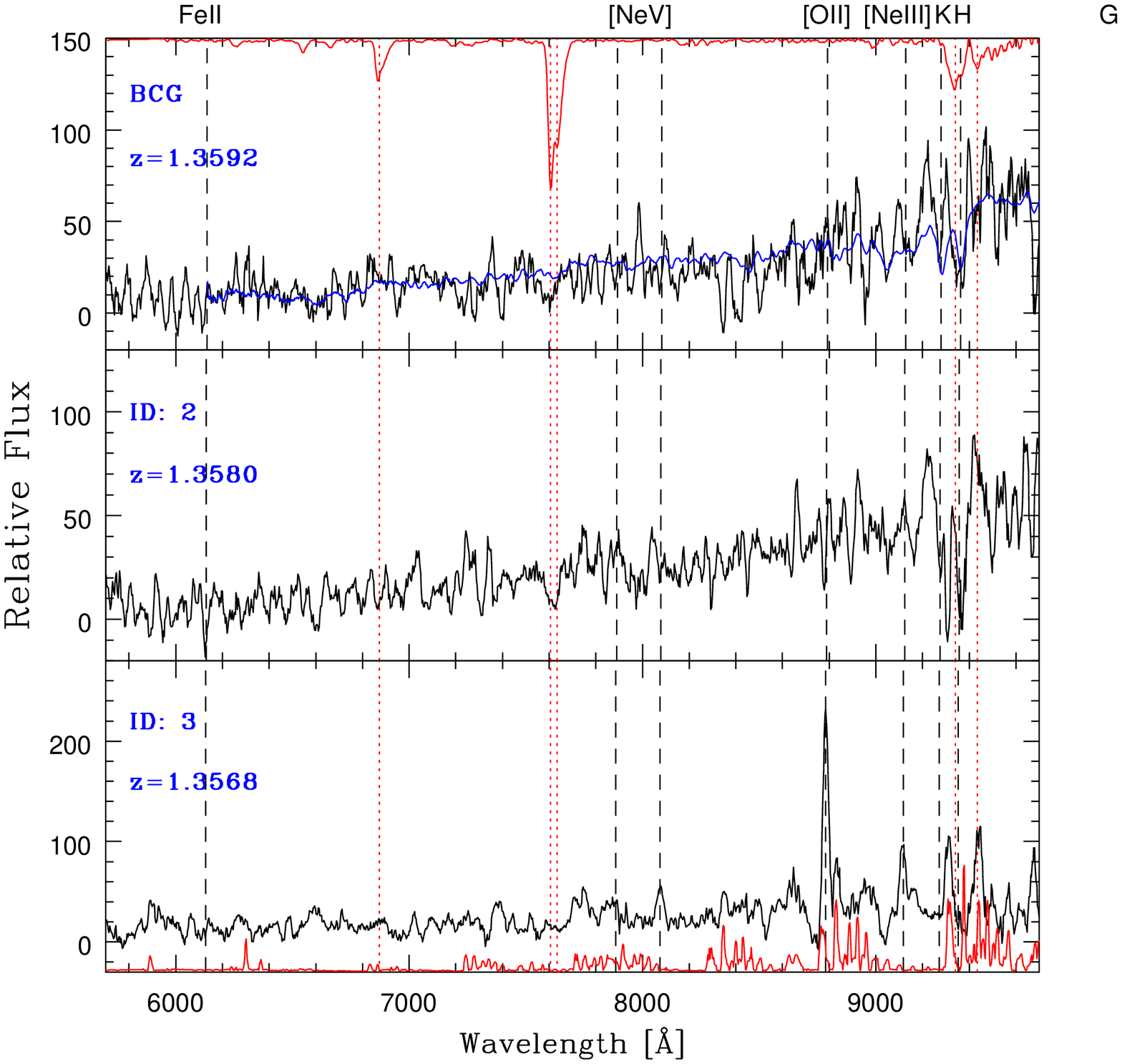} \\
\includegraphics[width=0.48\textwidth]{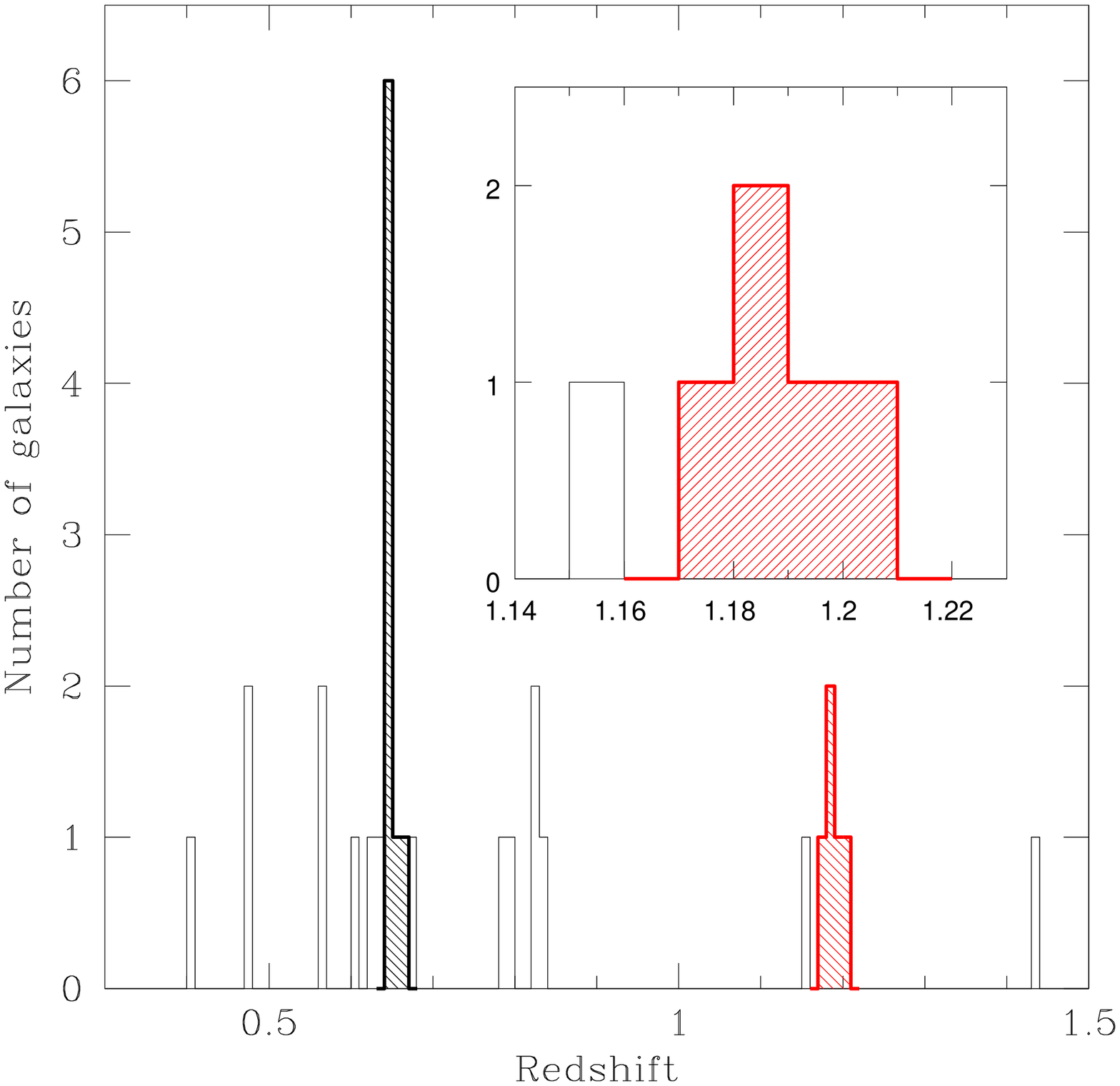}
\includegraphics[width=0.48\textwidth]{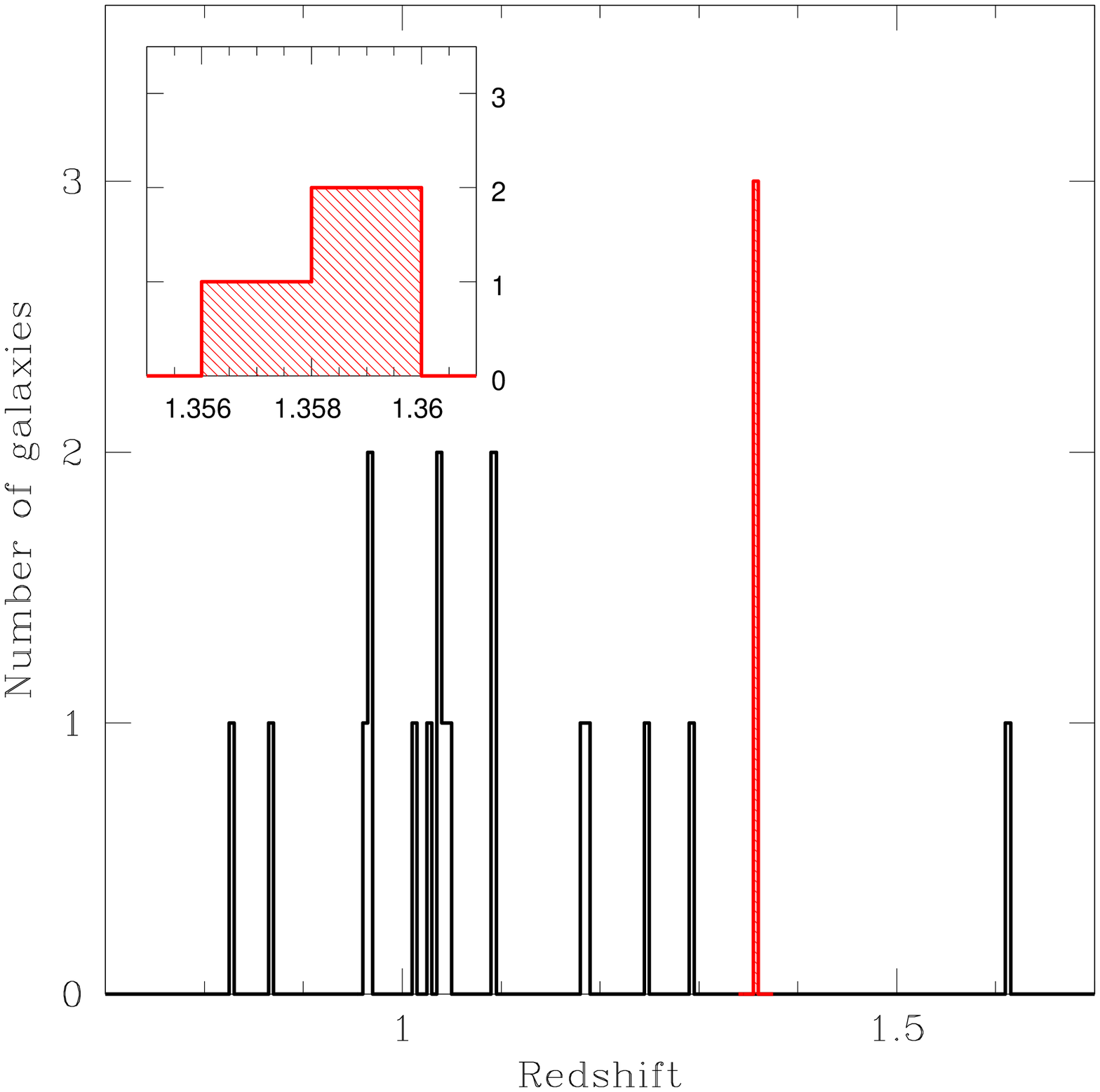}
\end{center}
\caption{\emph{Top:} Spectra of secure cluster members for \xa\ (left, $z=1.185$) and \xb\ 
(right, $z=1.358$) smoothed
with a 7 pixel boxcar filter. The expected observed positions of prominent spectral
features at the median redshift are indicated by black dashed lines. The sky spectrum
(bottom) and telluric features (top) are overplotted in red. For the BCGs (top panels)
we display an overlaid LRG template spectrum in blue. \emph{Bottom:} Distribution of
VLT/FORS2 galaxy spectra for the two clusters. The red hashed parts show the distribution
of the red galaxies for both clusters (also shown in the insets), selected by requiring
 the redshift to be within $z_{CL}\pm0.015$. The black hashed peak in the left panel
 (at $z\approx0.65$)
 corresponds to cluster \xc\ (see Appendix~\ref{sec:064}). See also Sect.~\ref{sec:spec} for discussion.}
\label{fig:spec}
\end{figure*}

In order to increase the efficiency of spectroscopic follow-up, we  submit the X-ray
identified candidates to a pre-screening process  based on optical/near-infrared imaging.
This allows us to compile a candidate shortlist with very low contamination rate
\citep{fassbender-thesis}.

\xa\ was classified as a good high redshift cluster candidate based on its solid
X-ray detection and a very prominent ($\sim25\sigma$) overdensity of red galaxies.
This system was therefore selected as a prime candidate for spectroscopic follow-up.

For \xb, the depth of the available
X-ray observations allowed us to establish it only as tentative high-z cluster
candidate. However, the optical
images revealed four very red galaxies exactly coincident with the X-ray emission
peak and thus also this system was included among the cluster candidates for the
purpose of spectroscopic confirmation with the VLT/FORS2 instrument.

For each cluster we carried out
an observation in the MXU-mode (Mask eXchange Unit), whose $6.8\arcmin \times 6.8\arcmin$
field-of-view allows us to obtain a sufficient number of galaxy spectra with a single slit mask
centered on the cluster.

We reduced the observations with a new FORS 2 adaptation of the VIMOS \emph{Interactive Pipeline and
Graphical Interface} \citep[VIPGI, ][]{scodeggio05} which includes all the standard analysis steps -  bias
subtraction, flat field corrections, image stacking and extraction of background
subtracted 1D galaxy spectra.
The wavelength calibration is carried out using a Helium-Argon lamp reference line spectrum
(calibration uncertainty $<1$\AA). Details of the spectroscopy reduction pipeline will be given
in Nastasi et al., in prep. The final stacked spectra are corrected for the sensitivity function
of the FORS 2 instrument. We obtain the galaxy redshifts by cross-correlating their spectra
with a galaxy template library using the IRAF\footnote{\texttt{iraf.noao.edu}}
package RVSAO and the EZ software \citep[][respectively]{kurtz98, garilli10}.
The redshifts are thus determined by taking into account the shape of the continuum
and all possible features (even those with low individual detection significances)
rather than just a few most significant absorption/emission features.

\subsubsection*{\xa}
\label{sec:064spec}
As can be seen in Fig.~\ref{fig:spec} (bottom, left), there is a peak of
six concordant redshifts in the galaxy redshift
distribution around the X-ray center of \xa\ at $z\approx1.19$.
 The measured redshifts, the most dominant spectral features
and cluster centric distances are listed in Table~\ref{tab:gals}.
Five of the galaxies are within $55\arcsec$ from the X-ray center.
This includes also the BCG and two additional red galaxies in the immediate vicinity of the
X-ray centroid ($<10\arcsec$ offset, see Fig.~\ref{fig:opt}), which allows us to establish
the redshift of the system with good confidence. The spectra of galaxies ID: 4 and ID: 5 have
low signal-to-noise ratios, but we still are able to measure their redshifts rather safely
and keep them therefore in the member list. Galaxy ID: 6 is not passive
- it exhibits an extremely
strong [O\,\textsc{ii}] emission line.

In order to estimate the final cluster redshift we apply a selection criterion
adopted from \citet{milvang-jensen08}: we require the galaxy redshifts to be in
a 0.015 wide redshift slice around the iteratively established cluster redshift.
This selection includes galaxies ID: $1-5$ and yields a median cluster redshift
of $z = 1.185 \pm 0.016$ (error is the interquartile range).

We conclude the discussion of the spectroscopy of \xa\ by remarking that
there is an additional spectroscopic galaxy redshift from the Sloan Digital Sky Survey which
is concordant with the cluster redshift. We discuss this source in Sect.~\ref{sec:ned}.

\subsubsection*{\xb}
For \xb\ we were able to obtain three galaxy spectra (Fig.~\ref{fig:opt},
top right). All the three spectra have good signal-to-noise ratios and the
redshifts can be anchored by several prominent spectral features
(Table~\ref{tab:gals}). They yield a redshift of the system equal to $1.358\pm0.001$.
The three spectroscopic
members are within $\sim6\arcsec$ from the X-ray center (Fig.~\ref{fig:opt},
bottom right). The fourth red galaxy close to the center was found to be a
foreground object.

Galaxy ID: 3 has a very prominent [O\,\textsc{ii}] emission line and
[Ne\,\textsc{iii}] and [Ne\,\textsc{v}] lines were detected as well. These features are
characteristic for the population of obscured AGN \citep[e.g.][]{groves06}.
The implications of the presence of an AGN for the X-ray analysis of the source is discussed
in Sect.~\ref{sec:physpar}.

Finally, we note that the redshift distribution in Fig.~\ref{fig:opt}
exhibits also a peak at $z\approx1.1$ (four concordant galaxy redshifts).
These galaxies are, however, spread over the whole field (i.e. are spatially unrelated)
and thus do not form a genuine system.

\subsection{Growth curve analysis of the X-ray imaging data}
\label{sec:gca}
\begin{figure*}[ht!]
\begin{center}
\includegraphics[width=0.48\textwidth]{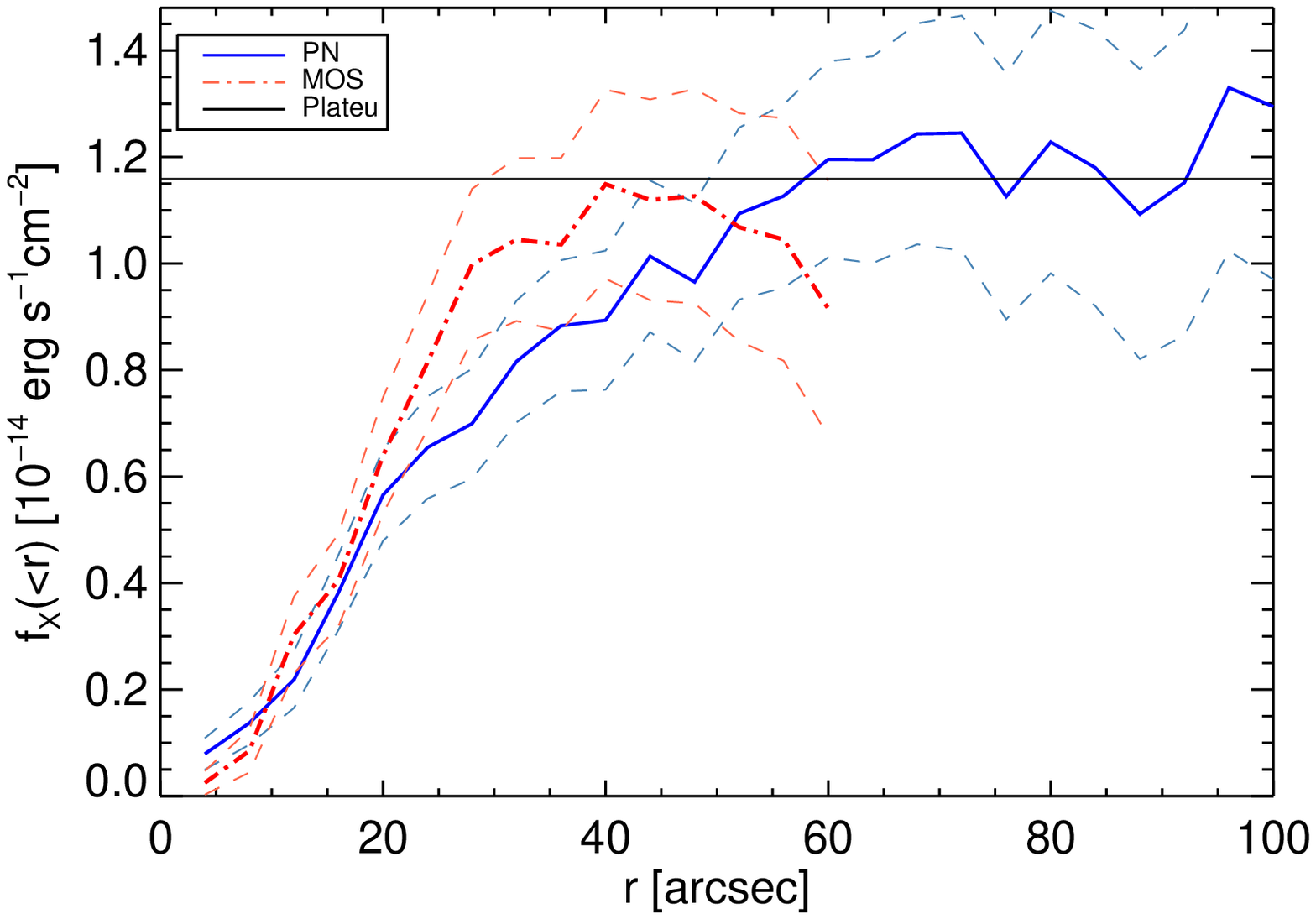}
\includegraphics[width=0.48\textwidth]{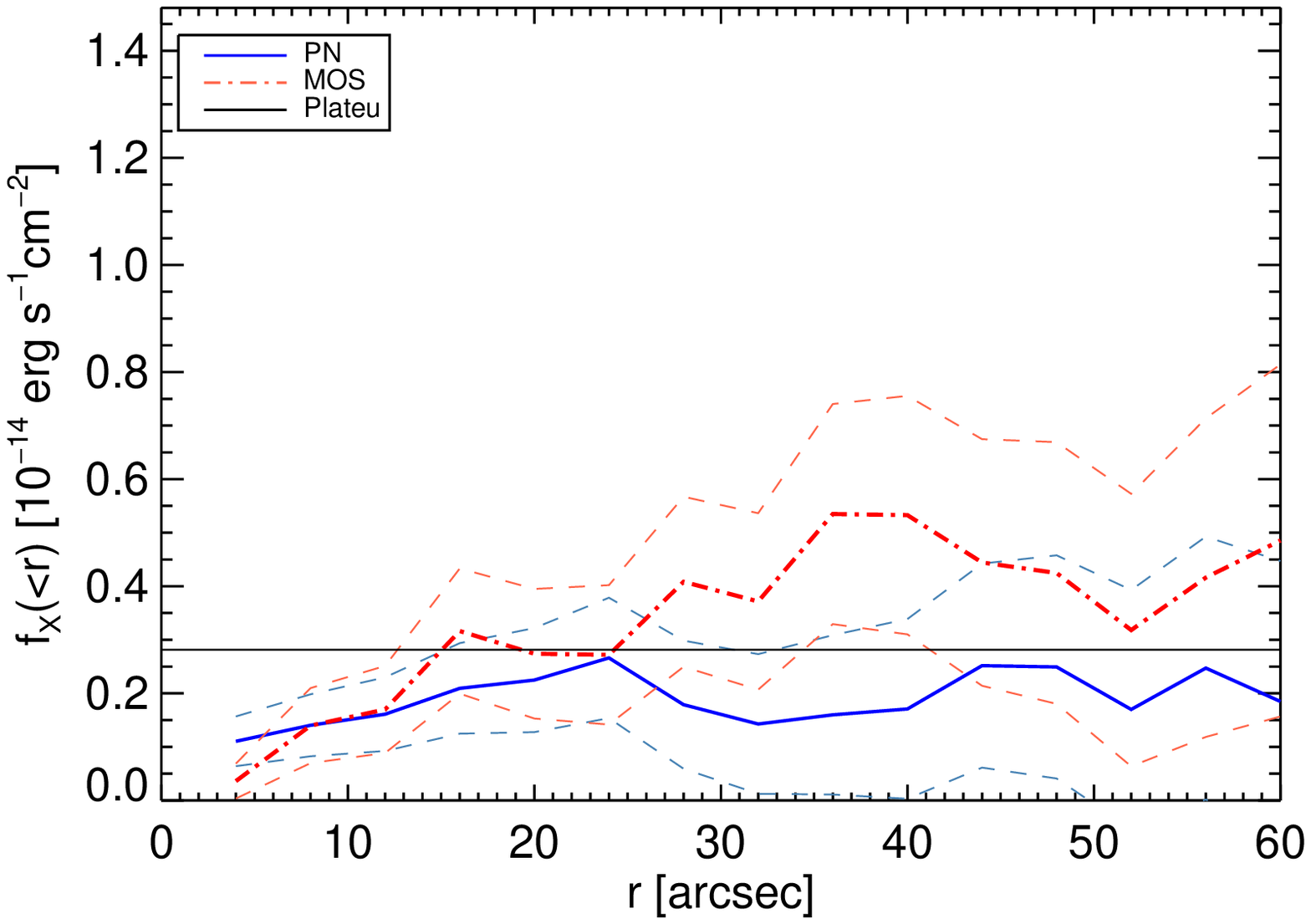}
\end{center}
\caption{Growth curve analysis of
\xa\ (z=1.185, left) and \xb\ (z=1.358, right). The curves show the encircled
cumulative
flux as a function of radius (PN: blue curve, combined MOS: red, dot-dashed).
Dashed lines mark the flux measurement error bars which include the
Poisson noise and an additional 5\% systematic error from the
background estimation. The horizontal lines mark the plateau
levels. See Sect.~\ref{sec:gca} for details.}
\label{fig:gca}
\end{figure*}

\begin{table*}
\begin{center}
\caption{Basic X-ray parameters of \xa\ and \xb. Flux and luminosity errors
 include the Poisson errors and a 5\% systematic error in the background
estimation.
 Errors of parameters obtained from scaling relations include
the measurement errors of the luminosity and temperature, respectively, and the
 intrinsic scatter of the scaling relations. The first set of
parameters uses the self-similar L$-$T and L$-$M relations from \citet{pratt09}
 for the bolometric luminosity. The second set of parameters (the no-evolution
case) follows the prescription of \citet{fassbender11a} by removing one $E(z)$
factor from the bolometric luminosity scaling relations of \citet{pratt09}.
In both scenarios, scaling relations are obtained by the BCES orthogonal fit
algorithm \citep{bces} and the input luminosities include the core regions.
Parameters for \xb\ are upper limits. See Sect.~\ref{sec:physpar} for more discussion.
}
\label{tab:physpar}
\centering
\begin{tabular}{ l c c l}
\hline
\hline
\textbf{Parameter} & \textbf{\xa} & \textbf{\xb}$^{a}$ & \textbf{Units} \\
\hline
 $\alpha$ (J2000)$^b$ & \,\,\,\,$03^{\mathrm{h}}\,02^{\mathrm{m}}\,11.9^{\mathrm{s}}$ & \,\,\,\,$15^{\mathrm{h}}\,32^{\mathrm{m}}\,13.0^{\mathrm{s}}$ \\
 $\delta$ (J2000)$^b$ & $-00\degr\,01\arcmin\,\,\,34.3\arcsec$ & $-08\degr\,36\arcmin\,\,\,56.9\arcsec$ \\
%%%%%%%%%%%%%%%%%%%%%%%%%%%%%%%%%%%%%%%%%%%%%%%%%%%%%%%%%%%%%%%%%%%%%%%%%%%%%
redshift & $1.185\pm0.016$ & $1.358\pm0.001$ &\\
E(z) & 1.96 & 2.15 &\\
Ang. scale & 8.28 & 8.41 & kpc s$^{-1}$ \\
n$_{\mathrm{H}}^a$ & 7.07 & 8.13 & $10^{20}$ cm$^{-2}$ \\
\hline
\textbf{L-M, L-T self-similar evol.} &&&\\
\hline
F$_{500}  $ [$0.5-2.0$ keV] & $11.44 \pm 1.28 $     & $3.02 \pm 0.96 $ & $10^{-15}$          erg cm$^{-2}$ s$^{-1}$ \\
L$_{500}  $ [$0.5-2.0$ keV] & $ 8.35 \pm 0.93 $     & $3.81 \pm 1.21 $ & $ 10^{43}$           erg s$^{-1}$           \\
F$_{500}  $ [bolometric] & $13.62 \pm 2.96 $        & $3.52 \pm 2.46 $ & $10^{-15}$ erg cm$^{-2}$ s$^{-1}$\\
L$_{500}  $ [bolometric]& $21.08 \pm 2.36 $         & $8.63 \pm 2.75 $  &      $ 10^{43}$           erg s$^{-1}$           \\
T$_{500}  $             & $2.8 \pm 0.7$             & $2.1 \pm 0.6 $   &       $        $          keV                    \\
M$_{500}  $             & $1.2 \pm 0.2 $           & $0.7 \pm 0.2  $  & $ 10^{14}$             M$_{\sun}$             \\
r$_{500}  $             & 0.47                      & 0.37             & Mpc \\
M$_{200}  $             & $1.7 \pm 0.4 $           & $1.0 \pm 0.3  $  & $ 10^{14}$             M$_{\sun}$             \\
\hline
\textbf{L-M, L-T no evol.} &&&\\
\hline
F$_{500}  $ [$0.5-2.0$ keV] & $11.59 \pm 1.33 $     & $2.89 \pm 1.14 $ & $10^{-15}$          erg cm$^{-2}$ s$^{-1}$ \\
L$_{500}  $ [$0.5-2.0$ keV] & $ 7.80 \pm 0.90 $     & $3.23 \pm 1.27 $ & $ 10^{43}$           erg s$^{-1}$           \\
F$_{500}  $ [bolometric] & $12.93 \pm 3.14 $ & $3.51 \pm 2.72 $ & $10^{-15}$ erg cm$^{-2}$ s$^{-1}$\\
L$_{500}  $ [bolometric]& $21.62 \pm 2.50 $     & $7.82 \pm 3.03 $  &      $ 10^{43}$           erg s$^{-1}$           \\
T$_{500}  $             & $3.5 \pm 0.9$         & $2.6 \pm 0.7 $    &       $        $          keV                    \\
M$_{500}  $             & $1.6 \pm 0.3 $       & $0.9 \pm 0.3  $   &    $ 10^{14}$             M$_{\sun}$             \\
r$_{500}  $             & 0.53                      & 0.42             & Mpc \\
M$_{200}  $             & $2.5 \pm 0.5 $       & $1.4 \pm 0.4  $   &    $ 10^{14}$             M$_{\sun}$             \\
\hline
\end{tabular}
\end{center}
\footnotesize{
$^a$~ All values are upper limits.
$^b$~X-ray coordinates based on a maximum-likelihood fit of a PSF-folded
beta model to the surface brightness distribution; $^c$ Values from the LAB HI survey
\citep{lab}.}
\end{table*}

We utilize an improved version of the \emph{growth curve method}
\citep[][]{boehringer00}, in order to trace the emission to an as high
cluster-centric distance as possible and obtain a reliable measurement of the flux. The cumulative source
flux (i.e. background-subtracted) as a function of radius, the growth curves, for the
two systems are displayed in Fig.~\ref{fig:gca}. The total source flux was determined
iteratively by fitting a line to the flat part of the background-subtracted growth curve.
We define the \emph{plateau radius} (r$_{\mathrm{plat}}$) as the aperture where
the growth curve levels off into a flat plateau. The flux at this radius is the total detected flux of the source.

\subsubsection*{\xa}
For \xa\ we found r$_{\mathrm{plat}}=55\arcsec$, F$_{\mathrm{plat}}(0.5 -
2.0$~keV$) = (11.73 \pm 1.36) \times 10^{-15}$ erg s$^{-1}$ cm$^{-2}$  and
a total luminosity L$_{\mathrm{plat}}(0.5 - 2.0$~keV$) =~(8.56~\pm~0.99)~\times~10^{43}$~erg~s$^{-1}$
(Fig.~\ref{fig:gca}).
Errors of the flux and luminosity include the Poisson errors and a 5\%
systematic error in the background estimation.

The analysis of this cluster is complicated by the presence of an extremely bright
point source $80\arcsec$ away from the cluster center. At this high off-axis
angle the point-spread function (PSF) is already significantly broadened with
respect to its on-axis shape and therefore the emission of the point
source is spread out in the PSF wings inside the cluster region. Before the background estimation we manually
removed a circular region with a conservative radius centered on the
point source. The flux estimation described above is based on
images with the point source masked out in the same way. We have mitigated this
contamination of the cluster emission by excluding these regions and extrapolating
the cluster emission as measured at the same cluster-centric radius but from the
uncontaminated parts.

\subsubsection*{\xb}
We display the growth curve for \xb\ in Fig.~\ref{fig:gca} (right)
extracted from the uncontaminated field (OBSID: 0100240701, see Sect.~\ref{sec:detect}).
We detected the source emission
 out to r$_{\mathrm{plat}}=22\arcsec$, with a total source flux of
F$_{\mathrm{plat}} (0.5 - 2.0$~keV$) = (2.82 \pm 1.11) \times 10^{-15}$ erg s$^{-1}$ cm$^{-2}$,
i.e. four times fainter than \xa.  This flux at the cluster's redshift
corresponds to a total luminosity L$_{\mathrm{plat}}(0.5 - 2.0$~keV$)
=~(3.59~\pm~1.41)~\times~10^{43}$~erg~s$^{-1}$. We note that the source is
very faint and thus the flux could be established only with a $\sim40\%$ error.
A systematic shift between the MOS and PN is also apparent in
Fig.~\ref{fig:gca} (the MOS flux being higher). Note however, that the growth
curves are cumulative flux distributions and thus the radial bins are not independent.
All subsequent physical parameters are thus only tentative and will require
a dedicated deeper (on-axis) X-ray observation for corroboration.

Except for pointing 0100240701 there is an additional XMM-\emph{Newton} observation available of
similar depth and at similar off-axis angle, but this one is heavily contaminated by
soft-protons. There is some evidence that the soft
protons undergo reflection on the telescopes mirror and are thus vignetted across the
field of view. The vignetting function for MOS was tentatively established by \citet{kuntz08} to
be shallower than the vignetting of genuine X-ray photons. For PN a systematic study has not
yet been carried out but a preliminary analysis suggests a similar shape to the MOS vignetting.
Given this, our two-component background model should be able to capture the
enhanced background to the first order. We have therefore extracted growth
curves also from the second observation. We find that while there is a relatively
large scatter between
the curves, they typically agree within the error bars. The measured total flux (the
plateau level of the curves) agree very well, the difference between them being much
smaller than their errors: F$_{\mathrm{plat}} (0100240701) = 2.82 \pm 1.11$ and
F$_{\mathrm{plat}} (0100240801) = 2.95 \pm 0.94$ in $10^{-15}$ erg s$^{-1}$ cm$^{-2}$
units in the $0.5 - 2.0$~keV band.
This is a reassuring indication that indeed the plateau fitting algorithm and
the procedure of combining the growth curves from PN and MOS yields very stable results,
even for observations with non-standard backgrounds.

\section{Results and discussion}
\label{sec:discuss}
\subsection{Physical properties of the clusters}
\label{sec:physpar}
The number of counts for both clusters is insufficient for a spectroscopic
analysis and therefore we can estimate additional physical parameters only through
luminosity based scaling relations. In the following analysis, we will assume
that the entire X-ray emission
detected and characterized by the growth curve analysis (Sect.~\ref{sec:gca})
originates in the ICM (after removing the detected point sources).
For \xa\  we find no indication that there is any further contamination,
but in the case of \xb\ this assumption is likely not valid due to the
presence of an obscured AGN (Sect.~\ref{sec:spec}). We discuss
this possibility in Sect.~\ref{sec:198discuss}. The physical parameters
for this system should therefore be considered as upper limits.

Due to the high redshift of the clusters and
the limited resolution of XMM-\emph{Newton} we can not excise the core regions of the
clusters. We therefore use the scaling relations that include cores.
We use the growth curve \citep{boehringer00} to iteratively obtain a
self-consistent set of parameters utilizing luminosity based scaling relations.
 The iterative procedure is described in \citet{suhada10}, with
the difference that we extrapolate the $0.5-2$~keV luminosity to obtain
 its bolometric value. We estimate the remaining physical parameters using
the bolometric L$-$T and L$-$M relations from \citet{pratt09}
(orthogonal fit, no Malmquist bias correction).

First, we assume a self-similar evolution of the
 scaling relations. Under this assumption, we estimate the objects to be
intermediate mass systems with M$_{500}\simeq 1.2 \times 10^{14}$ M$_{\odot}$
for \xa\ and M$_{500} \simeq 7 \times 10^{13}$ M$_{\odot}$ for \xb. This corresponds
to temperatures T$\simeq 3$~keV and T$\simeq 2$~keV respectively. The results for both
systems are summarized in Table~\ref{tab:physpar}.

The major uncertainty on the estimated physical parameters at these
high redshifts stems from the fact that the evolution of the scaling relations
is not yet well established. Self-similar evolution is a common assumption and
a direct prediction of the simple, purely gravitation driven growth. However there
are several indications that the evolution of luminosity scaling relations is
slower than the self-similar prediction - see discussion in \citet{fassbender11a}
and references therein, e.g. \citet{stanek10} and Reichert et al. (submitted).

We therefore adopt the simplified
approach of \citet{fassbender11a} and remove a factor of $E(z)$ from the
self-similar evolution factor ($E(z)^{-7/3}$) in the bolometric luminosity
based scaling relations. This modified evolution factor, $E(z)^{-4/3}$, is
consistent with preliminary results of Reichert et al. (submitted)
based on a fit to a large sample of high redshift clusters compiled from
the literature. Since our algorithm iteratively
estimates a self-consistent set of parameters, the change of scaling relations
impacts slightly the estimated flux and luminosity (mainly through the temperature
dependence of the energy-conversion-factor). This change is minuscule, but for
consistency we display the full sets of the estimated parameters for both
calculations (i.e. the self-similar and no-evolution scenarios) in Table~\ref{tab:physpar}.

However, the impact of the different evolution models on temperatures and masses
is serious. The non-evolving case
approach yields roughly $30\%$ higher cluster masses and $\sim25\%$ higher
temperatures (Table~\ref{tab:physpar}, bottom). Given the precision of the scaling
relations (and their intrinsic scatter) and the error of the luminosity measurement,
the estimates are still in agreement within their $1\sigma$ error bars (albeit for
the masses only barely). However, this uncertainty is systematic and very
important for studies of cluster samples (and naturally the eventual cosmological
constraints derived from them). This clearly demonstrates the importance of
establishing a well controlled high-redshift calibration cluster sample.

For both the self-similar and no-evolution scenarios we also include our estimates
of the mass M$_{200}$ (mass inside the aperture where the mean density is 200 times
the critical density of the Universe). The masses have been obtained by extrapolating
M$_{500}$ assuming an NFW profile \citep{navarro97} and using the relations of \citet{hukravtsov03}
and the DM profile concentration mass/redshift dependence of \citet{bullock01}.
The parameters are obtained iteratively using the M$_{500}$ values as inputs.
For both clusters the conversion factor M$_{500}\mapsto$M$_{200}$ is $\approx1.5$
and M$_{200}$ is $\sim90\%$ of the virial mass.

While at their observed redshifts the clusters would rank among intermediate
and low mass systems, respectively, they still have $\sim8-9$~Gyr of potential
mass accretion ahead, before reaching the current epoch. In order to predict
the final mass of the clusters at $z=0$, we use the mean mass growth rate
relations of \citet{fakhouri10}, based on the mass assembly histories of
halos in the Millennium and Millennium-II simulations. We estimate the
$z=0$ mass of \xa\ to be $1 \times 10^{15}$~M$_{\odot}$ and
$7 \times 10^{14}$~M$_{\odot}$ for \xb. Thus at the present epoch,
\xa\ would be a very massive clusters with a mass similar to the
Coma cluster. If we use the definition of formation time as the
redshift at which the cluster acquired $50\%$ of its $z=0$ mass
\citep[e.g. see the appendix of][]{giocoli07}, the formation redshift of \xa\ would
be around $z\approx0.5$, while \xb\ would be assembled slightly
earlier, at $z\approx0.6$.

\subsection{The nature of the X-ray emission of \xb}
\label{sec:198discuss}

\xb\ with its flux $F_{\mathrm{X}} \approx 3 \times 10^{-15}$~erg~cm$^{-2}$~s$^{-1}$
is one of the faintest cluster candidates discovered in a serendipitous X-ray survey.
Given the estimated upper limits, we are indeed entering here the low-mass cluster/group
regime at high redshifts. Probing the feasibility limits of this kind of cluster surveys,
however also means that we have to deal with increasing uncertainty in the sources'
classification and characterization.

In this case, the initial detection revealed the presence of an extended source at
$\sim2\sigma$ significance level (but only in one of the two observations).
Optical/near-infrared imaging confirmed the presence of red galaxies coincident
with the X-ray detection and spectroscopic data confirmed the presence of a dynamically
bound galaxy system.

Optical spectroscopy, however, also revealed the likely presence of an obscured AGN
in the core of the cluster (Sect.~\ref{sec:spec}). The X-ray spectral distribution
of an AGN can in first approximation be described as a power-law (with average index
$\Gamma\approx1.8$) intrinsically absorbed with hydrogen column densities from
$\sim10^{22}$~cm$^{-2}$ to over 10$^{25}$~cm$^{-2}$ (for Compton thick sources)
depending on the structure and orientation of the circumnuclear absorber
\citep[e.g.][]{antonucci85}. This local absorption introduces a photoelectric
absorption cut-off removing most of the soft X-ray emission. Unfortunately,
for an AGN at redshift $z=1.358$ a significant fraction of photons are redshifted
from unabsorbed parts of the spectrum into our detection band ($0.5-2$~keV).
Intrinsic absorption column densities equal to a few~times~$10^{23}$~cm$^{-2}$
are enough to remove a significant fraction of the soft emission even after
redshifting and thus for these cases the AGN contamination of the observed
X-ray emission should be small or even negligible. However, for lower column
densities the AGN emission would give a significant contribution and indeed
possibly be even the dominant source of the detected photons.

Since the observations are not deep enough to constrain the source spectrum,
we have checked the hardness ratios (ratio of the difference of counts in two
adjacent bands) of the source in our detection bands ($0.3-0.5$, $0.5-2$ and
$2-4.5$~keV). Due to the faintness of the source the ratios are highly uncertain,
but are consistent with ICM emission. Given the uncertainties, however, AGN
contamination (assuming moderate absorption) can also not be ruled out this way.

Apart from the spectral distribution, a safe detection of source extent would
constitute a strong piece of evidence that the observed emission originates
from the thermal bremsstrahlung of the ICM. As discussed in Sect.~\ref{sec:detect}
source extent was detected only in the slightly deeper, but contaminated field. We
re-examine the shallower, clean observation 0100240702 looking for instrumental
effects that could mask the source's true extent. In this observation the source
lies partially in a very prominent out-of-time (OoT) event stripe (in the PN
detector) caused by the very bright star system (UZ Lib) which was the actual
target of the observation. Originally, we removed the OoT stripe in the observation
in a standard, statistical way.\footnote{See e.g. the XMM-\emph{Newton} user
handbook, \texttt{xmm.esac.esa.int/sas/current/doc/}.} We then try an
alternative approach, by keeping the OoT events in the detection images and
modelling them in the background estimation step. This method also does not
yield an extent detection.

The area around the system is strongly affected by chip gaps in the MOS
detectors. In the next run we therefore applied a much less conservative
criterion for including low exposure areas in the vicinity of chip gaps,
gaining thus more geometric area for source detection. With this modification
the source is still detected without a significant extent.

As a final test, we carry out a joint source detection on both fields
simultaneously (i.e. two times three detectors, each in three bands).
The joint detection is carried out in two different ways. First we stack
the data from the same detectors/bands and run source detection simultaneously
on the nine merged data sets. \xb\ is detected in the merged data
set with a higher detection significance as in either of the single observations,
but its extent is not confirmed.

Merging observations has the disadvantage that the exact information on the shape
of the point-spread-function (PSF) is lost (the two observations have slightly
different off-axis and position angles. While the effect is expected to be small,
it could be a deciding factor in this case, since the potentially extended
emission is so weak. Therefore we also attempt to carry out source detections
on all 18 images\footnote{Two observations times three detectors times three
bands.} simultaneously without stacking them. There is currently no SAS task
that can carry out extended source detection in two pointings simultaneously,
but we modified the source code of the \texttt{emosaicproc} task (experimental
task in SAS v10.0.0,  originally developed for point source detection in mosaic
observations) to fit our purposes. However, also this approach does not yield
a detection of extent at a statistically significant level. We repeated the
procedures for several possible combinations of detectors with no extent
detections in most of the cases. Extent was detected exclusively in runs
where the two MOS detectors from the contaminated observations were used
along with either of the PN cameras. Given that these MOS detectors have
the highest contamination ($\sim90\%$ compared to $50\%$ of the PN from
this same observation) it leads us to the conclusion that the extent is
likely due to an unaccounted background fluctuations caused by the soft-proton
contamination. We estimate that for a fully conclusive confirmation of AGN
presence a $25$~ks \emph{Chandra} observation will be sufficient (gathering
$\sim25$ source counts). Time for this follow-up observation has been
allocated.

In summary, the extent of the source was not confirmed by deeper analysis.
While \xb\ clearly constitutes a dynamically bound system, the detected
X-ray emission can not be unambiguously attributed to the ICM from the
available data. Note however, that we have below 100 source counts
(after background subtraction) even if we combine both available
observations. Given our findings we can not exclude the possibility
that the detected AGN is the dominant (possibly only) source of
X-ray emission detected from this system. This source is thus an example
of systems that even with availability of multi-wavelength data
are hard to properly classify. For large upcoming surveys such
systems will be presumably numerous (e.g. \emph{eRosita}, which in
addition has a slightly worse PSF) and therefore additional
studies will be needed to establish how can we explore the cluster-group
transition regime at high redshifts with good confidence and effective
use of follow-up observing time.

\subsection{The galaxy population of the clusters}

It is interesting to note that in the cluster \xa\ we detect [O\,\textsc{ii}]
line emission in four out five spectroscopic members (excluding the galaxy ID: 6).
This feature, an indicator of sustained star-formation, is detected along with other
features which are typical for mature systems.
Similar activity is observed also in other high redshift X-ray selected clusters:
XMMU~J1007.4+1237 \citep[at $z=1.56$][]{fassbender11b}, XMMU~J0338.8-0030
(Pierini et al., submitted) and XMMU~J2235.3-2557 \citep{lidman08}
at $z=1.39$. While some of these [O\,\textsc{ii}] emitters are bluer
than the cluster red sequence, many of them have colors in full agreement
with the old, passive galaxies and in some cases are even redder (e.g.
in XMMU~J2235.3-2557). These galaxies can also span a large range in
magnitudes, up to the very bright end of the cluster luminosity function.
An increase in star formation activity in red sequence galaxies is also
apparent in optically selected cluster sample of \citet{finn10} (mostly
intermediate redshift systems) and in dense galaxy environments at redshifts
$\sim1$ seen in GOODS and DEEP2 galaxy surveys \citep{elbaz07,gerke07,cooper08}.

We are thus very likely observing residual stochastic star-formation in probably
bulge dominated disc galaxies. This effect can be expected to be more important
as we move to higher redshifts and enter lower-mass regimes. As we have
remarked in Sect.~\ref{sec:physpar}, based on cluster mass growth rates
from simulations, \xa\ is still in its assembly phase and is expected
to be experiencing significant mass accretion and merging activity. These
processes lead to large variations in the cluster/group tidal field. Based
on numerical simulations \citet{martig08} show that tidal field fluctuation
can enhance the star formation activity of galaxies (beyond the expectations
from purely galaxy-galaxy interaction driven activity). This effect should be particularly
efficient at high redshifts and in low mass systems, before quenching processes
take place. The XDCP project has the X-ray sensitivity and sky area to be
able to effectively study this transition regime and the relevant environmental
effects. We leave further discussion to an upcoming dedicated study based on
available data.  However, we note that in order to completely disentangle the
ongoing processes, information on the spectral energy distribution at
$\lambda > 4\,000$~\AA~(rest frame)  is necessary as well as high-spatial
resolution imaging in order to be able to assess the galaxy morphologies.

\subsection{Cross-correlation with known sources}
\label{sec:ned}
We queried the NASA/IPAC Extragalactic Database\footnote{\texttt{nedwww.ipac.caltech.edu}}
and the SIMBAD Astronomical Database,\footnote{\texttt{simbad.u-strasbg.fr/simbad/}}
in search for potentially interesting known sources.

We find that \xa\ has been previously detected by the BLOX survey \citep[Bonn
lensing, optical, and X-ray selected galaxy clusters][]{dietrich07}
as the object BLOX~J0302.2-0001.5. The cluster was
selected in X-rays, but not by the optical and weak lensing
detection algorithms. Their estimates of the X-ray parameters
(r$_C=12.8\arcsec\pm1.2\arcsec$ and flux F$_{X} = (12.1 \pm 1.3)
\times 10^{-15}$ erg cm$^{-2}$ s$^{-1}$ in the $0.5-2$ KeV band) are
in good agreement with our values. The cluster does not have a
redshift measurement from the BLOX survey.

At a cluster centric distance of $\sim185\arcsec$ we find
the source SDSS~J030214.82+000125.3 identified in the
Sloan Digital Sky Survey (SDSS) as a quasar \citep{cetty06}.
The object has a known spectroscopic redshift, $z=1.179$ \citep{schneider07, cristiani96},
which is in concordance with our redshift for \xa. At $\sim185\arcsec$
offset (corresponding to  $\sim1.5$ Mpc at this redshift) it could
be associated with the cluster's outskirt region. We also detect this
quasar as a high significance X-ray point sources in our XMM-\emph{Newton}
observation (see Fig.~\ref{fig:xim}).

For the cluster \xb\ we do not find any complementary redshifts in the databases.
Neither do we find any known radio source within a $2\arcmin$ radius
from either system.

Both sources are also part of the The Second XMM-\emph{Newton} serendipitous
source catalog\footnote{\texttt{amwdb.u-strasbg.fr/2xmmidr3/catentries}}
\citep{watson09}. Their detection parameters are in agreement with our estimates,
however since our pipeline is optimized for high redshift cluster
detection, we detect both systems with slightly higher detection likelihoods.

\section{Conclusions}
\label{sec:conclusions}

\begin{enumerate}

\item We have detected two high redshift systems, \xa\ at $z=1.185$ and \xb\ at
$z=1.358$. The objects were X-ray selected in the framework of the XMM-\emph{Newton}
Distant Cluster Project.

\item We have obtained and analysed medium deep optical/near-infrared imaging and
deep optical spectroscopy with VLT/FORS2 and measured spectroscopic redshifts for
both systems. We have confirmed \xa\ as a bona fide galaxy cluster. Among its
spectroscopically confirmed members we find several luminous [O\,\textsc{ii}] emitters.
These giant galaxies are likely experiencing residual stochastic star formation activity,
possibly triggered by galaxy-galaxy interactions and fluctuations in the overall tidal field.

\item
Based on the obtained optical/near-infrared imaging we established that \xa\ 
has a well populated red sequence. In fact, \xa\ corresponds to one of the most
prominent overdensities of red galaxies ($1.45<\mathrm{z}-\mathrm{H}\leq2.15$)
among the known X-ray selected $z>1$
clusters. Given the currently limited depth of the data for \xb, we can see
the bright end of the red sequence (finding two spectroscopical members
to have colors consistent with a SSP spectro-photometric sequence for the cluster redshift),
but deeper observations will be required
to study the galaxy population of the cluster in more detail.

\item From archival XMM-\emph{Newton} data we have estimated the basic
physical parameters of \xa. Within the r$_{500}$ aperture we measured
the luminosity (0.5-2 keV band) of cluster to be
L$_{500} =  (8.35 \pm 0.93) \times  10^{43}$ erg s$^{-1}$. Assuming
a self-similar evolution of the L$-$M scaling relation this value
correspond to M$_{500} = (1.2 \pm 0.2) \times 10^{14}$~M$_{\sun}$.
This ranks \xa\  among intermediate mass clusters at its redshift.

\item We confirm the presence of a dynamically bound galaxy system
with three concordant redshifts and coincident with \xa. We also find
[O\,\textsc{ii}], [Ne\,\textsc{iii}] and [Ne\,\textsc{v}] emission lines in the
optical spectrum of one of the member galaxies making it a
likely obscured AGN candidate.

\item
We carried out an in-depth X-ray analysis of \xb, showing that its
original tentative detection as an extended source can not be confirmed
by currently available data.
While the nature of the X-ray emission as originating from faint ICM
emission can not be ruled out, we conclude that it is likely that the emission is dominated
(or possibly even fully caused) by the central AGN.
Notwithstanding this, we estimate upper limits on the X-ray properties
for the case that the AGN emission is negligible.
We estimate the upper limit for  the $0.5-2$~keV band luminosity to be
L$_{500} =  (3.81 \pm 1.21) \times  10^{43}$~erg~s$^{-1}$ and the corresponding
mass M$_{500} = (0.7 \pm 0.2) \times 10^{14}$~M$_{\sun}$.

\item We have discussed the effect of non-self similar evolution of the scaling
relations on our mass estimates. We find that a no-evolution scenario yields
up to $30\%$ higher masses and $\sim25\%$ higher temperatures at these redshifts.
This strongly underscores the importance of the efforts to properly calibrate
these relationships in the redshift regime z$\gtrsim0.8$.

\item We detected and analysed a third cluster, \xc, which was
serendipitously detected together with cluster \xa.
This cluster is established to be an intermediate mass system
at an intermediate redshift, $z=0.647$.

\end{enumerate}

We are experiencing a time period when many crucial questions about the cluster
population and its evolution can start to be addressed by analysing cluster
samples at high redshifts. This is also the
objective of the XDCP project, with the main aim to address the evolution of
scaling relations and obtain cosmological constraints. The present paper
extends the XDCP sample and provides the first analysis of
the clusters in preparation for planned deeper studies based on additional
optical/near-infrared data. \xb\ is scheduled for deep J and Ks imaging
by the Hawk-I instrument on the VLT. A \emph{Chandra} observation to
investigate the nature of the X-ray emission from this system has
also been allocated.
For \xa, K band imaging has already been
obtained by the Large Binocular Telescope.
A joint, multi-wavelength analysis of these (and other XDCP) targets will be discussed
in up-coming studies.

\begin{acknowledgements}
We acknowledge the support provided by the VLT staff in
carrying out the service observations. This work was supported by the DFG under
grants Schw536/24-1, Schw 536/24-2, BO 702/16- 3, and the German DLR under
grant 50 QR 0802. RS acknowledges support by the DFG in the program SPP1177.
HB acknowledges support for the research group through
The Cluster of Excellence 'Origin and Structure of the Universe',
funded by the Excellence Initiative of the Federal Government of
Germany, EXC project number 153.
DP acknowledges the kind hospitality of the Max-Planck-Institute
for extraterrestrial Physik.
HQ thanks the FONDAP Centro de Astrofisica for
partial support.
The XMM-\emph{Newton} project is an ESA Science Mission with instruments and
contributions
directly funded by ESA Member States and the USA (NASA). This research has made
use of the NASA/IPAC Extragalactic Database (NED) which is operated by the Jet
Propulsion Laboratory, California Institute of Technology, under contract with
the National Aeronautics and Space Administration. We thank Andreas Reichert
for providing his cluster compilation catalog. We are thankful to Rosario
Gonzales-Riestra, Marcella Brusa, Mara Salvato and Angela Bongiorno
for useful discussions. We thank the anonymous referee for a fast
reply and helpful suggestions.
\end{acknowledgements}

\bibliographystyle{aa}
\bibliography{clusters}

%%%%%%%%%%%%%%%%%%%%%%%%%%%%%%%%%%%%%%%%%%%%%%%%%%%%%%%%%%%%%%%%%%%
\begin{appendix}

\section{\xc}
\label{sec:064}

The cluster \xc\ was detected in the observation OBSID: 0041170102
only $\sim2\arcmin$ from \xa\ at an off-axis angle of $11\arcmin$,
with a high confidence extent significance ($\sim10\sigma$).

We measured the redshift of the cluster (z=$0.647\pm0.003$)
from the same VLT/FORS2 data taken for \xa. The redshift is
based on the \citet{milvang-jensen08} criterion
(Sect.~\ref{sec:spec}), identifying 8 cluster members (black
hashed peak in Fig.~\ref{fig:spec}, bottom left).

\begin{figure}[ht!]
\begin{center}
\includegraphics[width=0.5\textwidth]{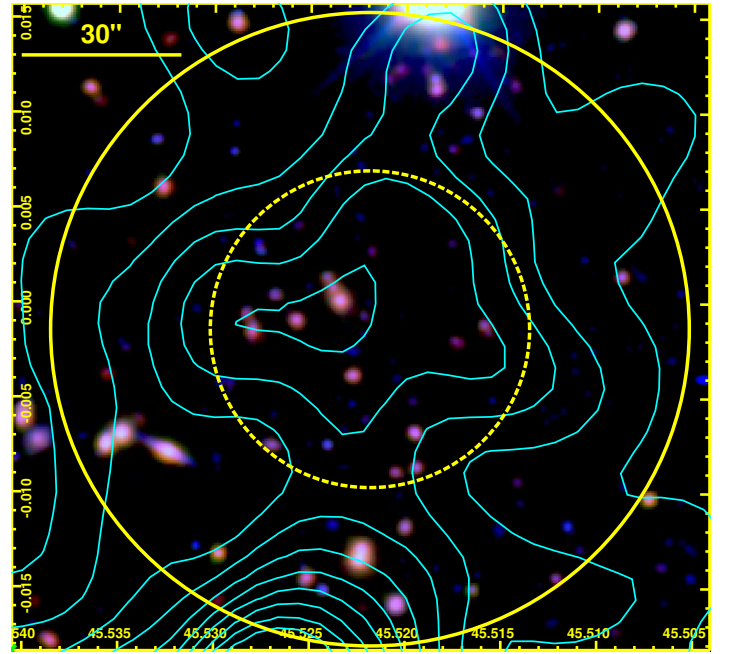}
\end{center}
\caption{Pseudo-color image of the clusters \xc\ (left, red channel: H band,
green: z, blue: R). Adaptively smoothed X-ray contours are overlaid in cyan.
Solid/dashed circles mark a $60\arcsec$/$30\arcsec$ radius region centered on
the X-ray detection.}
\label{fig:opt064}
\end{figure}

This redshift places the cluster below the formal XDCP distant cluster sample
limit (z$\ge0.8$). The galaxy distribution of the cluster
well matches the X-ray surface brightness distribution
(see Fig.~\ref{fig:opt064}), with the BCG close to the
X-ray peak.

\begin{figure}[ht!]
\begin{center}
\includegraphics[width=0.5\textwidth]{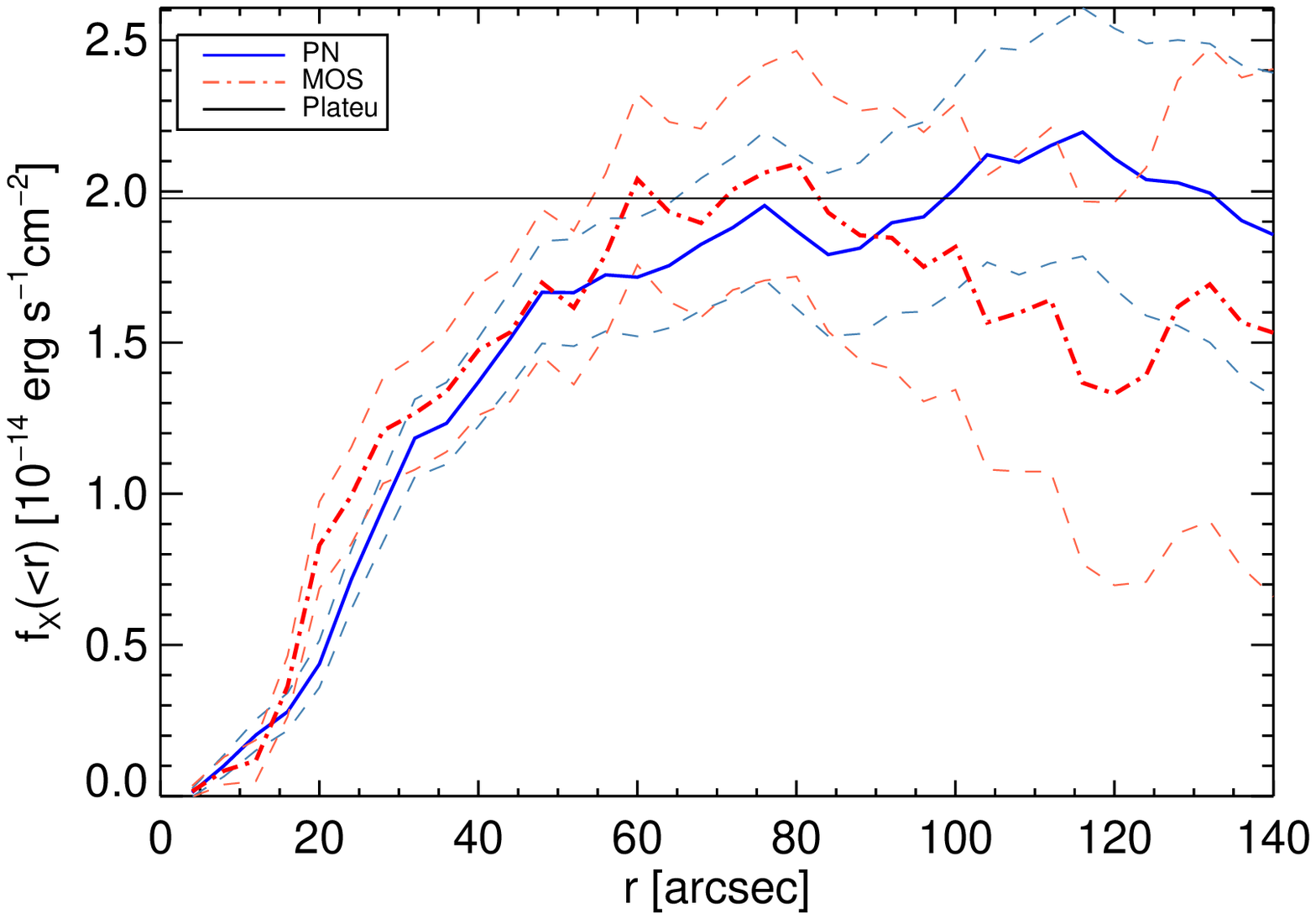}
\end{center}
\caption{Growth curve analysis of
\xc. The curves show the encircled
cumulative flux as a function of radius (PN: blue curve, combined MOS: red, dot-dashed).
Dashed line marks the flux measurement error bars which include the
Poisson noise and an additional 5\% systematic error from the
background estimation. The horizontal line marks the plateau
level.}
\label{fig:gca064}
\end{figure}

We carry out the X-ray analysis as delineated in
Sect.~\ref{sec:detect} and Sect.~\ref{sec:gca}.
In a $1\arcmin$ aperture we detect 140 source counts in PN and 114 in MOS.
The estimated beta model core radius is $r_{C}=30.9\arcsec$.
The growth curves are displayed in Fig.~\ref{fig:gca064}. Both PN
and MOS curves are in good agreement and have well established
plateau levels with F$_{plat} = (19.83 \pm 2.16) \times 10^{-15}$
erg cm$^{-2}$ s$^{-1}$.

We estimate the cluster's mass to be M$_{500} = (1.0 \pm 0.2) \times
10^{14}$ M$_{\odot}$ from its measured luminosity
L$_{500} = (3.26 \pm 0.33) \times 10^{43}$ erg s$^{-1}$ ($0.5-2$~keV). This
corresponds to a $2.3$~keV temperature. The effect of the evolution
uncertainty described in Sect.~\ref{sec:physpar} is slightly smaller
than for the $z>1$ cases - the no-evolution scenario yields a $\sim20\%$
higher M$_{500}$ and $\sim13\%$ higher temperature. The physical parameters
are summarized in Table~\ref{tab:physpar064}.

\begin{figure}[ht!]
\begin{center}
\includegraphics[width=0.5\textwidth]{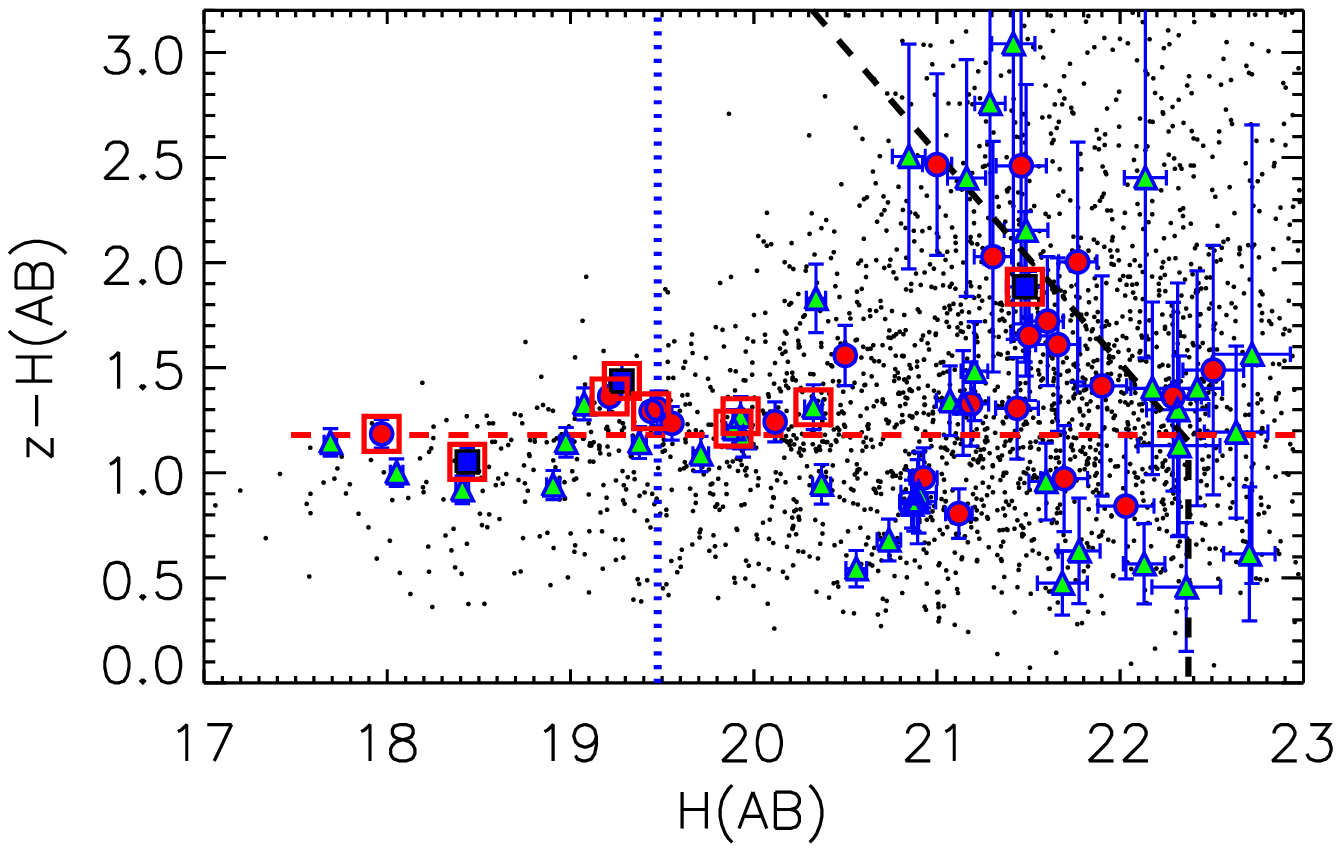}
\end{center}
\caption{The z$-$H vs. H color-magnitude diagram of the \xc\ cluster's field.
Red boxes mark secure spectroscopic cluster members. Galaxies with
projected cluster centric distance less than $30\arcsec$ are shown as
red circles, with distances between $30\arcsec-60\arcsec$ as green triangles.
Galaxies with concordant redshift at $>60\arcsec$ distances have blue squares.
The
dashed black line marks the $50\%$ completeness limit.
The apparent H band magnitude of an L$^{*}$ galaxy at cluster redshift
($z=0.647$) is shown with a vertical blue dotted line. We overplot the z$-$
H color of a solar
metallicity SSP model with formation redshift z$_f=5$ and
age $\sim6.3$~Gyr (corresponding to the cluster redshift) as a reference
(horizontal red dashed line).
The presence of a red sequence with analogous colors is evident.}
\label{fig:cmd064}
\end{figure}

In Fig.~\ref{fig:opt064} we display the pseudo-color image of \xc\ 
and in Fig.~\ref{fig:cmd064}. The individual frames and photometry
are the same as for \xa\ and described in Sect.~\ref{sec:spec}.

The CMD can be found in Fig.~\ref{fig:cmd064}.
The cluster has a rich red sequence with a BCG coincident with the
X-ray centroid. Although the BCG seems to experience
a merging activity, its color is in agreement with the SSP model
prediction for the cluster redshift (red dashed line in Fig.~\ref{fig:cmd064}).

\begin{figure}[ht!]
\begin{center}
\includegraphics[width=0.5\textwidth]{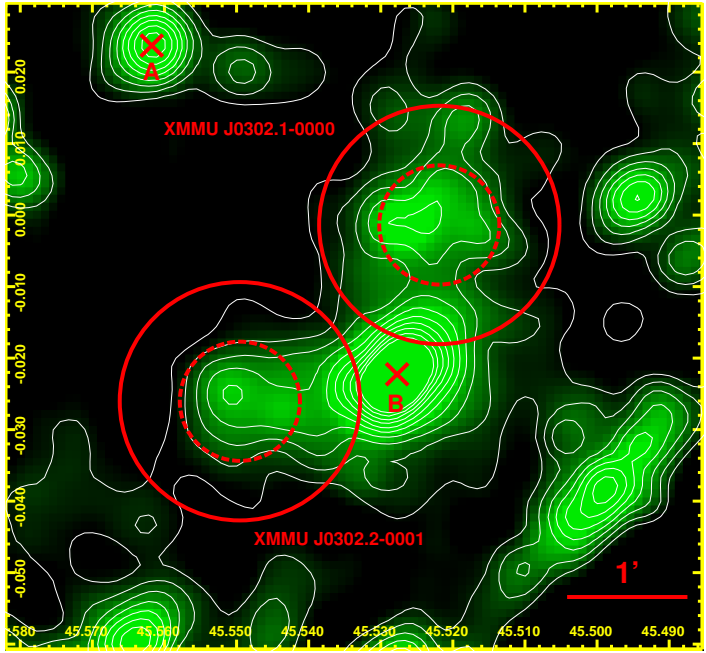}
\end{center}
\caption{Adaptively smoothed $0.5-2$ keV X-ray image of the wider
neighborhood of \xa\ and \xc. The red circles have $60\arcsec$
radii, while the dashed circles $30\arcsec$. X-ray contours
 are overlaid in white. Point A
marks the AGN at the redshift of \xa\ (SDSS J030214.82+000125.3)
and point B the very X-ray bright AGN SDSS~J030206.76-000121.3, that
 has a concordant redshift with the cluster \xc.}
\label{fig:xim}
\end{figure}

We have obtained spectroscopy also for the very bright X-ray AGN
$\sim75\arcsec$ from the cluster center (see Fig.~\ref{fig:xim})
and find that its redshift is concordant with \xc. At the cluster
redshift this is equivalent to $\sim0.5$~Mpc, i.e. the AGN is
associated with the cluster.
This source is also
contained in the SDSS catalog \citep{schneider07} (quasar
SDSS~J030206.76-000121.3 at redshift $z=0.641$).

Similarly to \xa, the cluster \xc\ is also part of the BLOX survey catalog
\citep[see Sect.~\ref{sec:ned} and ][]{dietrich07}, detected
independently in the X-ray data and through an optical matched filter
cluster finder (object ID: BLOX~J0302.0-0000.0). The estimated
X-ray extent of $r_{C}=27.8\arcsec \pm 1.9\arcsec$ is fully
consistent with our value.
The $0.5-2$ keV flux estimated by the BLOX survey
is F$_{X}=(29.4 \pm 2.6) \times 10^{-15}$~erg cm$^{-2}$ s$^{-1}$,
i.e. significantly higher than our value.
We note however, that their estimate is based on the direct output
of the detection pipeline, whereas ours is from a dedicated growth
curve analysis which includes several improvements: 1) a visual
screening and manual adjustment of masks for contaminating
sources (indeed there is a bright point source detected only
$20\arcsec$ from the cluster's core); 2) the redshift and
temperature dependence of the energy conversion factor
(which is needed to convert the detected counts to flux)
is implemented in an iterative fashion and 3) we use the proper
response file calculated locally for the clusters position.
Finally, the largest part of the difference comes from the fact that
the \citet{dietrich07} flux is extrapolated to
infinity (assuming a beta-model), while our estimate corresponds
to a (finite) aperture and is model independent. For a cluster with
a large core radius ($\sim31\arcsec$), there is a comparatively larger
fraction of the total flux (extrapolated to infinity) outside r$_{500}$
than for a cluster with smaller core radius (such as \xa\ 
($r_{\mathrm{C}}=14\arcsec$), where the agreement with the BLOX survey
value is much better).

The BLOX survey significantly underestimates the cluster redshift
(their value is 0.4). They estimate a cluster richness of
$\lambda_{CL} = 65.9$, where $\lambda_{CL}$ is the equivalent number
of L$^{*}$ galaxies with the same total optical luminosity
as the cluster galaxies \citep[for the exact definition see][]{postman96}.

The cluster is also part of the Second XMM-\emph{Newton} serendipitous
source catalog \citep{watson09}. The source parameters in this
catalog are in good agreement with our values.

In the NED database we find within a $4\arcmin$ query radius four galaxies
with photometric redshifts in agreement with the cluster redshift
(i.e. in the range $0.6-0.7$) from \citet{waskett04}. This includes
the already mentioned SDSS~J030206.76-000121.3.

In summary, \xc\ is confirmed as an intermediate mass system at
intermediate redshift. While it is below the redshift threshold of the
XDCP distant cluster sample, owing to its proximity to \xa\ it will
benefit from upcoming deeper multi-wavelength follow-up data and
will be thus an interesting object in its own right.

\begin{table}
\begin{center}
\caption{Basic X-ray parameters of \xc. See Table~\ref{tab:physpar}
for explanations.}
\label{tab:physpar064}
\centering
\begin{tabular}{ l c l}
\hline
\hline
\textbf{Parameter} &  & \textbf{Units} \\ %\textbf{\xc}
\hline
 $\alpha$ (J2000)$^a$ & \,\,\,\,$03^{\mathrm{h}}\,02^{\mathrm{m}}\,05.3^{\mathrm{s}}$  \\
 $\delta$ (J2000)$^a$ & $-00\degr\,00\arcmin\,\,\,05.0\arcsec$  \\
%%%%%%%%%%%%%%%%%%%%%%%%%%%%%%%%%%%%%%%%%%%%%%%%%%%%%%%%%%%%%%%%%%%%%%%%%%%%%
redshift & $0.647\pm0.003$  &\\
E(z) & 1.43 &\\
Ang. scale &6.92& kpc s$^{-1}$ \\
n$_{\mathrm{H}}^b$ & 7.05 & $10^{20}$ cm$^{-2}$ \\
\hline
\textbf{L-M, L-T self-similar evol.} &&\\
\hline
F$_{500}  $ [$0.5-2.0$ keV] & $19.20 \pm 1.95 $ & $10^{-15}$ erg cm$^{-2}$ s$^{-1}$ \\
L$_{500}  $ [$0.5-2.0$ keV] & $ 3.26 \pm 0.33 $ & $10^{43}$ erg s$^{-1}$           \\
F$_{500}  $ [bolometric]    & $22.78 \pm 4.33$ & $10^{-15}$ erg cm$^{-2}$ s$^{-1}$ \\
L$_{500}  $ [bolometric]    & $7.71 \pm 0.78 $     & $10^{43}$ erg s$^{-1}$ \\
T$_{500}  $                 & $2.3 \pm 0.7$         & keV                    \\
M$_{500}  $                 & $1.0 \pm 0.2 $       & $ 10^{14}$ M$_{\sun}$  \\
r$_{500}  $                 & 0.56                 & Mpc/arcsec \\
M$_{200}  $                 & $1.4 \pm 0.3 $       & $ 10^{14}$ M$_{\sun}$  \\
\hline
\textbf{L-M, L-T no evol.} &&\\
\hline
% 101210b summary_064_0041170102.log
F$_{500}  $ [$0.5-2.0$ keV] & $18.88 \pm 2.01 $ & $10^{-15}$ erg cm$^{-2}$ s$^{-1}$ \\
L$_{500}  $ [$0.5-2.0$ keV] & $ 3.14 \pm 0.33 $ & $10^{43}$ erg s$^{-1}$           \\
F$_{500}  $ [bolometric] & $21.19 \pm 4.56$ & $10^{-15}$ erg cm$^{-2}$ s$^{-1}$ \\
L$_{500}  $ [bolometric]    & $7.73 \pm 0.82 $     & $10^{43}$ erg s$^{-1}$ \\
T$_{500}  $             & $2.6 \pm 0.6$             & keV                    \\
M$_{500}  $             & $1.2 \pm 0.3 $           & $10^{14}$ M$_{\sun}$   \\
r$_{500}  $             & 0.59             & Mpc \\
M$_{200}  $             & $1.7 \pm 0.4 $           & $10^{14}$ M$_{\sun}$   \\
\hline
\end{tabular}
\end{center}
\footnotesize{$^a$~X-ray coordinates based on a maximum-likelihood fit of a PSF-folded
beta model to the surface brightness distribution; $^b$ Values from the LAB HI survey
\citep{lab}.}
\end{table}

\clearpage
\section{High redshift cluster detections in the past decade}
\label{sec:hzhist}
We have argued in Sect.~\ref{sec:intro}, that in the recent years
much important progress has been made in the gradual construction
of statistically large, high redshift cluster samples.
These samples will allow us to calibrate the
scaling relations to redshift $\approx1$ and beyond and
constrain evolutionary models for the ICM and the clusters'
galaxy populations.

\begin{figure}[ht!]
\begin{center}
\includegraphics[width=0.5\textwidth]{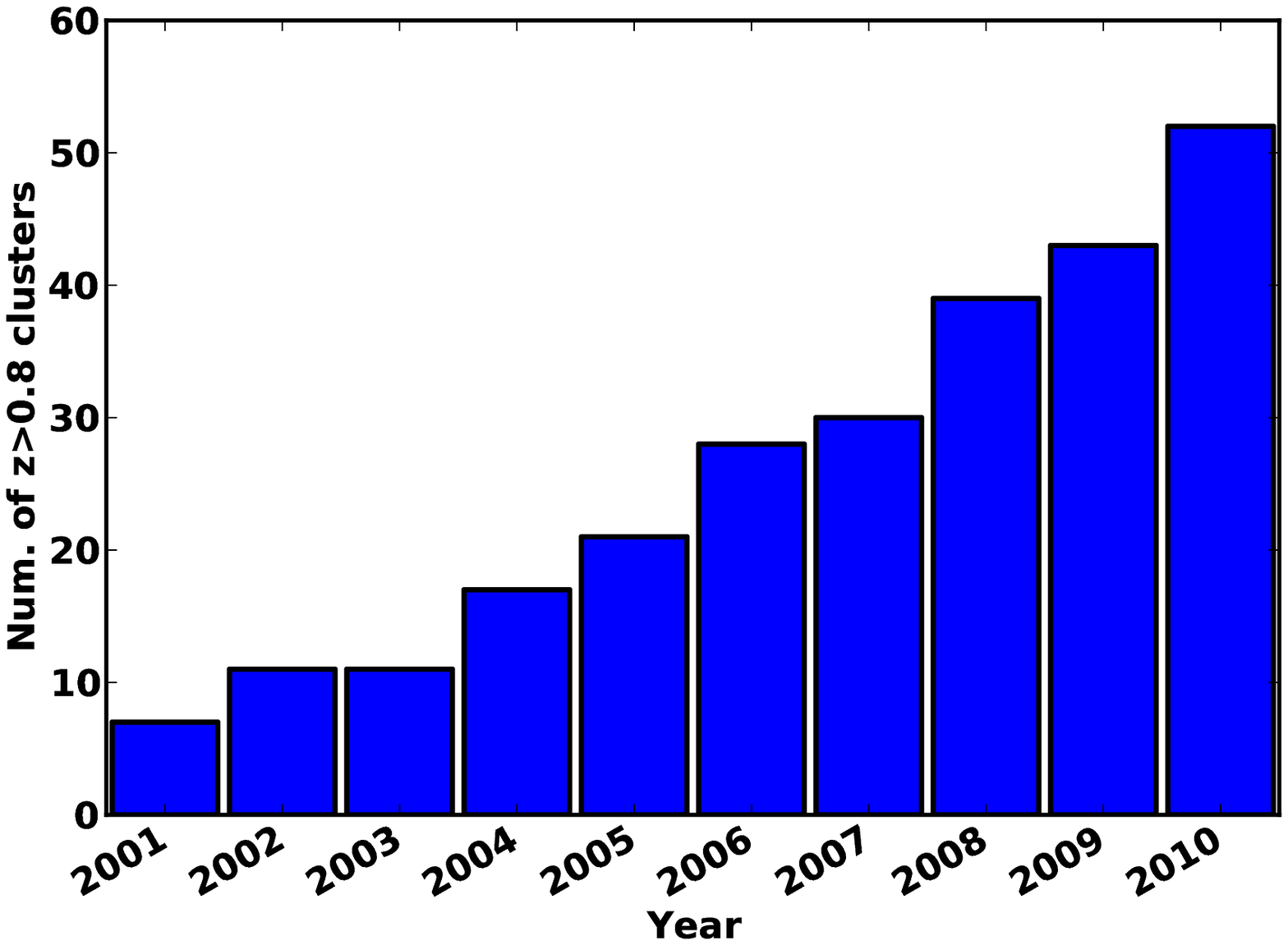}
\includegraphics[width=0.5\textwidth]{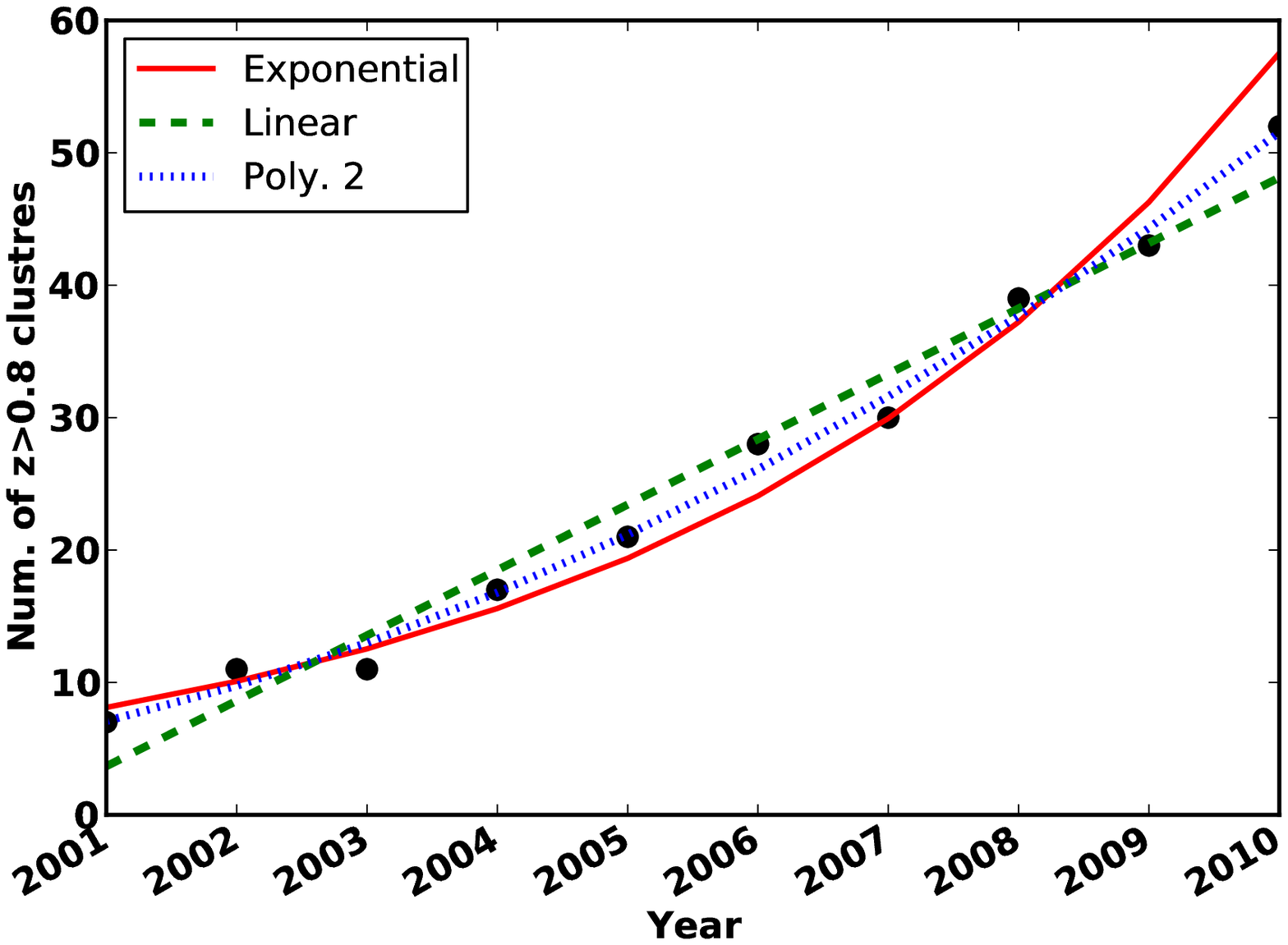}
\end{center}
\caption{\emph{Top:} Cumulative histogram of the known clusters
with spectroscopic redshift  $>0.8$ and an X-ray luminosity measurement
compiled by Reichert et al. (submitted). \emph{Bottom:} Linear (green dashed line),
exponential (red full line) and second order polynomial (blue dotted line) fits to the
data in the top panel. See the text for discussion.}
\label{fig:hzcumul}
\end{figure}

Reichert et al. (submitted) compiled a list of known clusters published
up to year 2010 (including), which have secure spectroscopic redshifts and
an X-ray luminosity measurement. We select from this catalog a high
redshift subsample based on the XDCP project's criterion, i.e. clusters
with redshifts $z>0.8$. While the aim of this catalog is to compile
clusters from larger samples, care was taken to include also individually
reported high redshift objects. The catalog utilises the latest analysis
of each cluster if several are available and therefore whenever
it is possible we replace this reference with the year of first \emph{spectroscopic redshift measurement}
(understood here as the discovery year). We aim here just for a simple qualitative
analysis and therefore these minor effects do not influence our conclusions.
The cumulative histogram of the compiled catalog of  $z>0.8$ clusters is
shown in Fig.~\ref{fig:hzcumul}. As we can see the progress made in the past
decade (2001-2010) is truly impressive. The total number of clusters given
our criteria is 52. Only six of these clusters were known before the year 2001.

In the bottom panel of Fig.~\ref{fig:hzcumul} we check whether the
total number of clusters grows linearly with time (green dashed line).
While the fit is acceptable, there is an indication that the last few years the detection
rate has been even larger. An exponential relation\footnote{An
exponential growth is motivated as a potential instance of
Moore's law. A similar growth is observed not only in
improvement of computing hardware (and other digital electronic devices),
but given a suitable figure-of-merit also in several scientific subfields,
e.g. the total number of particles in cosmological N-body simulations
or the number of DNA base pairs sequenced per year.
}
 yields an equally good fit (in the sense of the R$^{2}$ statistic).
An exponential growth might be also preferred if we relax the criteria and would
include also clusters with only photometric redshift estimates and not
having X-ray luminosity measurements. Especially the
Sunyaev-Zel'dovich surveys are currently (e.g. since $2009$)
the main purveyors of cluster samples with high median redshifts \citep[e.g. $\sim0.6$
from the South Pole Telescope survey,][]{vanderlinde10}.
 We also overplot a second order polynomial in Fig.~\ref{fig:hzcumul} (blue dotted line)
which well describes the observed detection counts (and confirms the preference
for accelerating detection rates). The prediction of this model is
59 clusters given our selection criteria by the end of year 2011.

Upcoming large area X-ray surveys (with XMM-\emph{Newton} and eventually
\emph{eRosita}) along with other cluster selection approaches (SZE, optical)
and the new near-infrared spectrographs will enable us
to further increase our high-z cluster samples. Independently of the
exact shape of the growth of our cluster catalogs,
 the future of high redshift cluster studies certainly seems very promising.

\end{appendix}
\end{document}